\documentclass[11pt,hyper]{JHEP3}
\usepackage{epsfig}
\usepackage{calc}
\def\lsim{\mathrel{\rlap{\lower4pt\hbox{\hskip1pt$\sim$}}
   \raise1pt\hbox{$<$}}}         
\def\gsim{\mathrel{\rlap{\lower4pt\hbox{\hskip1pt$\sim$}}
    \raise1pt\hbox{$>$}}}         
\newcommand{\bsg}{$b \rightarrow s \gamma$}

\newcommand{\ds}{{\sffamily DarkSUSY}}
\def\msun{M_{\odot}{\ }}
\newcommand{\code}[1]{{\tt #1}}

\newcommand{\bea}{\begin{eqnarray}}
\newcommand{\eea}{\end{eqnarray}}
\newcommand{\be}{\begin{equation}}
\newcommand{\ee}{\end{equation}}
\newcommand{\ba}{\begin{array}}
\newcommand{\ea}{\end{array}}

\title{Direct versus indirect detection in mSUGRA with self-consistent halo models}

\author{Joakim Edsj\"o, Mia Schelke\\ Department of Physics, AlbaNova, Stockholm University, SE-106 91 Stockholm, Sweden\\ 
E-mail: \email{edsjo@physto.se, schelke@physto.se}}
\author{Piero Ullio\\ SISSA, via Beirut 4, I-34014 Trieste, Italy\\
  E-mail: \email{ullio@sissa.it}}

\abstract{
We perform a detailed analysis of the detection prospects of neutralino dark matter in the mSUGRA framework. We focus on models with a thermal relic density, estimated with high accuracy using the \ds\ package, in the range favored by current precision cosmological measurements. Direct and indirect detection rates are computed implementing two models for the dark matter halo, tracing opposite regimes for the phase of baryon infall, with fully consistent density profiles and velocity distribution functions. This has allowed, for the first time, a fully consistent comparison between direct and indirect detection prospects.
We discuss all relevant regimes in the mSUGRA  parameter space, underlining relevant effects, and providing the basis for extending the discussion to alternative frameworks. In general, we find that direct detection and searches for antideuterons in the cosmic rays seems to be the most promising ways to search for neutralinos in these scenarios.
}

\begin{document}

\section{Introduction}

The identification of dark matter in the Universe is one of the most compelling
targets in Science today. In the ``concordance'' cosmological model~\cite{triang}, emerging from precision cosmological measurements and from tests of the theory of
structure formation, some unknown form of non-baryonic cold dark matter (CDM)
accounts for  about 30\% of the mean energy density of the Universe today. 
Among the solutions of the dark matter puzzle, the most natural scheme is the one in which CDM, analogously to the  baryonic and radiation components,  appears as a thermal relic from the early Universe; in particular, weakly interacting massive particles (WIMPs) are natural candidates, as their thermal relic abundance is automatically of the right order of magnitude. Moreover, it is a scheme with strong motivations from the particle physics point of view: the most widely studied WIMP dark matter candidate is the lightest neutralino in supersymmetric extensions of the Standard Model of particle physics.

Several techniques have been studied to search for dark matter WIMPs 
(for thorough reviews and comprehensive lists of references, see, e.g.~\cite{jkg,larsrev}). One of the issues that is often raised, eventually to understand in what direction experimental efforts should be focussed, regards the comparison between capabilities of different techniques. Actually, as we already mentioned, the idea of WIMPs is associated to a scheme rather than to a model, and no simple recipe can be given. It is only when the setup of a specific model  is fully defined that a sensible comparison can be performed; on general grounds, the only statement one can safely formulate is that, in most cases, different techniques probe different properties of WIMP dark matter (we refer here to properties related both to particle physics and astrophysics), hence they are complementary. On the other hand, it is true that focussing on a specific model is  sometimes a very useful exercise, which allows for better understanding of at least some aspects of an otherwise too complicated problem. In the context of supersymmetric extensions to the Standard Model, this is often done, e.g.,  to present limits from current sets of data,  or to address the reach of future accelerator experiments; the same is useful applied to dark matter searches.

In this paper we will focus on the supersymmetric neutralino as a dark matter candidate in the so-called mSUGRA framework of supersymmetry breaking, one of the simplest and most popular models. Direct and indirect detection of dark matter in this model has been the subject of several studies (a non-exhaustive list of recent papers on this topic includes, e.g., \cite{arnowitt-dir,matchev-ind,bottino-dir,ellis-dir,orloff-dir-nu,hooper-all,baer-dir-ind,donato-pbar}). 
Our reiteration starts from the point of view of restricting to configurations corresponding to relic abundances in the range currently favored by cosmological measurements; to fulfill  this requirement,  we calculate relic densities with the \ds\ computer code \cite{ds}, which includes all possible sfermion, neutralino and chargino coannihilations and is currently the most accurate code for relic density calculations. We will then consider two limiting cases of profiles for the dark matter halo of the Milky Way, consistent with available constraints, and derive self-consistent density and velocity distributions for the neutralinos in two such sample halo models. This approach will allow, for the first time, a fully consistent comparison of the prospects to detect relic neutralinos with direct detection or various indirect detection methods, like neutrinos from the Earth/Sun or cosmic rays from annihilations in the galactic halo. Our aim is also to try to make this comparison more transparent than in previous studies, and to underline step by step what the properties are that enter critically to make the balance bend on one side or another. This analysis will then try to clarify what the characteristics are of the model we are considering, and at the same time to help in foreseeing what may change in other setups.

The outline of the paper is as follows. We will start with an introduction to the supersymmetric framework, continue with a discussion of the self-consistent halo profiles we use, go through the various detection rates for models within the measured relic density range and finally end with a discussion and conclusions.

\section{The particle physics model}

\subsection{The supersymmetric setup}

The present analysis is performed in the mSUGRA setup~\cite{msugra}, 
namely the N=1 supersymmetric extension of the Standard Model (SM) 
defined in a supergravity
inspired framework and with the smallest possible number of free parameters:
on top of a structure with minimum field content, universality is
assumed at the grand unification (GUT) scale, both in the gaugino 
and the scalar sector of the theory.  The mSUGRA action is then fully defined, 
by only four parameters and one sign: the GUT scale values of the soft 
supersymmetry breaking fermionic mass parameter $m_{1/2}$,  
the scalar mass parameter $m_0$ and the trilinear scalar coupling $A_0$, 
the ratio of the vacuum expectation values of the two neutral components 
of the SU(2) Higgs doublets $\tan\beta$, and the sign of the Higgs superfield 
parameter $\mu$ (the absolute value of $\mu$ is fixed by electroweak 
symmetry breaking; regarding the convention on its sign, following, e.g.,~\cite{HK},
it is assumed here that $\mu$ appears with a minus sign in the superpotential).

The appropriate set of renormalization group equations (RGEs) allows to relate
univocally the GUT scale structure to the low energy (weak scale) spectrum of
the theory. Here, soft breaking parameters, gauge and Yukawa couplings are evolved
down to the weak scale with the ISASUGRA RGE code as given in version
7.67 of the ISAJET software package~\cite{isajet} (introducing some minor
changes, such as, the conversion of the ISASUGRA code to double precision
to improve on its stability; for more details on this and other technical points 
on the code implementation and on the interface with the \ds\ package,
see the discussion in~\cite{coann}).
The mSUGRA setup is probably the most popular framework for studying 
supersymmetric (SUSY) extensions of the Standard Model,  and its rather constrained low energy 
structure has been extensively discussed. We review here
very briefly those features which will be relevant in our discussion of the detection
prospects for SUSY dark matter in this framework. 

Our working hypothesis is that the lightest supersymmetric particle (LSP) (and 
our dark matter candidate) is the lightest neutralino, defined as the lightest mass
eigenstate from the superposition of the two neutral gaugino and the two neutral 
Higgsino fields:
\begin{equation}
  \tilde{\chi}^0_1 = 
  N_{11} \tilde{B} + N_{12} \tilde{W}^3 + 
  N_{13} \tilde{H}^0_1 + N_{14} \tilde{H}^0_2\;.
\end{equation}
The coefficients $N_{1j}$, obtained by diagonalizing the neutralino mass matrix, 
are mainly a function of the bino and the wino mass parameters $M_1$ and $M_2$, 
and of the parameter $\mu$. 
From the assumption of gaugino mass unification at the GUT scale, 
it follows that the hierarchy between $M_1$ and $M_2$ at the weak scale 
is fixed to about $M_1 \simeq 0.5 M_2$, and hence that the wino component of the 
LSP is always very small. There are then essentially just two regimes in the 
composition of the lightest neutralino: in most of the allowed regions in 
a generic scan in the $m_0$--$m_{1/2}$ parameter space, the neutralino LSP is a 
very pure bino; in a rather thin slice at $m_0 \gg m_{1/2}$, sometimes dubbed 
the ``focus point'' region~\cite{focus}, on the border with the region where there is 
no radiative electro-weak symmetry breaking, the parameter $\mu$ 
is driven to small values and forces a mixing between the gaugino and 
Higgsino sectors, with a considerable Higgsino fraction in the lightest neutralino. 

Regarding SUSY scalars, except at very large values of  $m_0$, i.e. for 
very heavy sfermions which do not play much of a role, the RGEs drive 
the slepton sector to be lighter than the squark sector: in particular in the 
cosmologically interesting regime $m_0 \lsim m_{1/2}$, the lightest 
stau is always the lightest sfermion, possibly even lighter than the 
lightest neutralino 
if $m_0 \ll m_{1/2}$; selectrons and smuons are slightly heavier, while the lightest
stop, which is also the lightest squark, is significantly heavier and scalar partners
of the light quarks, whose exchange can contribute to the scattering of 
neutralinos on protons and
neutrons, are even more massive. A notable exception to the pattern just described
is given by the case when the trilinear coupling, $A$, in the stop mass matrix is tuned to
such a value that one of the two stop mass eigenstates becomes light:
in this case, a stop can become the next-to-lightest SUSY particle or even the 
LSP, while the spectrum of other SUSY scalars is not significantly changed.

Finally, regarding the Higgs sector, the mass of the light CP-even Higgs 
boson $H^0_2$, especially for low  $\tan\beta$ and low $m_0$, 
is mainly set by $m_{1/2}$ and changes only slightly with
$m_0$, while the masses of the heavy CP-even Higgs boson $H^1_0$ and
of the CP-odd $H^3_0$ increases monotonically with both $m_0$ and  
$m_{1/2}$. In most regions of the parameter space $H^2_0$
decouples and behaves like the Standard Model Higgs; at large $\tan\beta$,
the mass of the heavy Higgs bosons can be driven to rather small values, 
and in a diagonal stripe of  the parameter space $m_{1/2}$-$m_0$, the mass 
of the LSP happens to match one half of the mass of the heavy 
Higgs boson(s), giving rise to resonant annihilations of neutralinos.

We will focus on a few selected scans in the mSUGRA parameter space,
encompassing however all these features, whose role on dark matter detection
we will enlighten.

\subsection{Constraints on the model}

The first pattern of model discrimination we apply is based on the calculation of the
relic abundance of the neutralino LSP for any given set of the parameters,
and by the requirement that the relic density matches
the best fit value from the latest cosmological measurements of the 
non-baryonic dark matter component in the Universe.

The relic density calculation we implement for this analysis is included in a 
new extended 
version of the \ds\ package \cite{ds}, which has been recently released. It is suitable
for a proper treatment of  any coannihilation effect, applying, at the same time, the 
state of the art technique to trace the freeze-out of a species in the early 
Universe~\cite{GondoloGelmini}, with a careful numerical treatment of resonance
and threshold effects, and full numerical solution of the density evolution equation
(avoiding approximations such as, e.g., the expansions in powers of the relative 
velocity that is often applied). This method is described in detail 
in Ref.~\cite{coann},
where it has been applied to the SUSY setup considered here; assuming masses, 
widths and couplings of particles in the model are given with an adequate precision,  
this neutralino relic abundance calculation has an estimated precision of 
1\% or better.

The combined analysis of the latest data on the cosmic microwave background 
ani\-sotropÑies and large scale galaxy surveys gives a fairly accurate estimate for
the cold dark matter contribution to the energy density of the Universe.
We take, as reference value, the best fit value derived in Ref.~\cite{SDSS+WMAP},
under standard assumptions: 
$\Omega_{CDM} h^2 = 0.103^{+0.020}_{-0.022}$ 
(Table 3, first column, in~\cite{SDSS+WMAP}), where $\Omega_{CDM}$ is 
the ratio between mean CDM density and critical density and $h$ is the Hubble 
constant in units of 100 km s$^{-1}$ Mpc$^{-1}$.

So far, experimental searches for SUSY particles have given null results.
To exclude models that violate accelerator constraints, we adopt 
the compilation of limits on SUSY masses as given by the Particle Data 
Group 2002  (PDG) \cite{pdg02}. On top of these, we consider two more
restrictive conditions: We assume as limit on the lightest chargino mass
the final kinematical limit of LEP, i.e. 103.5~GeV~\cite{chargino}; 
also, as we mentioned that
$H_2^0$ is in most case SM-like, we will use as a limit on the
$H_2^0$ mass, the current limit on the SM Higgs boson mass, 
114.1~GeV~\cite{SMHiggs}.

Another strong bound is given by the SUSY contributions to the rare 
decay \bsg. Our estimate of this process includes the complete next-to-leading 
order (NLO) correction for the SM contribution and the dominant
NLO corrections for the SUSY term.  The NLO QCD SM calculation is performed
following the analysis in Ref.~\cite{bsgsm}, modified according to~\cite{bsgmagic}, 
and gives a branching ratio 
$\mathrm{BR}[B\rightarrow X_s\,\gamma] =3.72\times10^{-4}$ 
for a photon energy greater than  $m_b/20$. In the SUSY contribution, we 
include the NLO contributions in the two Higgs doublet model,
following~\cite{bsgh2}, and the corrections due to SUSY particles. The latter
are calculated under the assumption of minimal flavour violation, with the 
dominant LO contributions from Ref.~\cite{bsgtan}, and with the NLO QCD term
with expressions of \cite{bsgsusy} modified in the large $\tan\beta$ regime
according to~\cite{bsgtan}. In the mSUGRA framework
(see, e.g., ~\cite{bsgcompare}), the largest discrepancy 
between the LO and the NLO SUSY corrections are found for ${\rm sign}{\mu}>0$, 
large $\tan\beta$ and low values of $m_{1/2}$: in this case the SUSY 
contribution to the decay rate is negative, and the discrimination of models 
based on the NLO analysis is less restrictive than the one in the LO analysis.
We will assume as allowed range of branching ratios 
$2.0\times10^{-4}\leq\mathrm{BR}[B\rightarrow X_s\,\gamma] \leq4.6\times10^{-4}$,
which is obtained adding a theoretical uncertainty of $\pm0.5\times10^{-4}$ to the
experimental value quoted by the Particle Data Group 2002~\cite{pdg02}. 

Finally, we consider the information on the model following from the SUSY 
contribution to the anomalous magnetic moment of the muon. Our estimate of the 
$(g-2)_\mu$ SUSY term is based on the analysis in Ref.~\cite{g-2cal}. 
This can be compared with the range of values favored by the latest BNL
measurements, relative to the calculation in the standard  model~\cite{g-2},
$-5 \times 10^{-10} < a^{\rm exp}_\mu - a^{\rm SM}_\mu < 47 \times 10^{-10}$
[$2\sigma$]. The Standard Model prediction depends on experimentally measured quantities that can be obtained from $\tau$ decay or $e^+e^-$ annihilation data. As these two methods give different results, we have, to be conservative, taken the lower limit derived from $\tau$ data and the upper limit from $e^+e^-$ data. This gives the rather large range of allowed $(g-2)_\mu$-values stated above. With this broad range of allowed values, it turns out that the $(g-2)_\mu$ does not place any tight constraints on the models we consider here. Hence, all models we show in the later sections are compatible with $(g-2)_\mu$.

\section{Models for the dark matter halo}

The distribution of dark matter particles in the halo of the
Galaxy plays a major role when making predictions for dark matter detection
rates. Unfortunately, that distribution is rather poorly known, and one
has to rely on large extrapolations. The approach we follow here is to take
N-body simulations of hierarchical clustering in cold dark matter 
cosmologies as a guideline. Numerical results indicate that dark matter 
halos can be described fairly accurately by a universal density profile of given 
outer slope and singular towards the galactic center 
(there is a general consensus on the $r^{-3}$ scaling at large radii, while there 
is still an on-going debate regarding how cuspy the profiles are), 
and by only two extra parameters, e.g.\ a length scale 
and the normalization of the density profile at that scale, or better, in a language 
which is more appropriate in a structure formation study, the virial mass of 
the object $M_{vir}$ and its concentration parameter $c_{vir}$ 
(roughly speaking, the total mass 
and some measure of what fraction of this mass is confined in the inner 
portion of the halo). Actually, since the first systematic studies~\cite{NFW},
it has emerged that $M_{vir}$ and $c_{vir}$ are, at a given redshift, highly
correlated, with smaller objects being more concentrated, an effect which has 
been understood in terms of the form of the power spectrum of density 
perturbations and of the redshift of collapse of a given class of 
objects~\cite{bullock,ens}.

N-body simulation results provide snapshots of halos before the baryon infall;
the formation of the luminous components of the Galaxy is likely to have 
induced a back-reaction on the structure of the dark halo as well.  We will 
consider two opposite and at the same time extreme regimes for this effect. 
In the first, the baryon infall occurs 
as a smooth and slow process, with no net transfer of angular momentum 
between baryonic and non-baryonic terms. This is the regime in which the dark
halo gets adiabatically contracted, a process in which, in the limit of spherical 
systems with unchanged local velocity distribution (as it happens, e.g., if all 
particles are placed on circular orbits), the mass distributions in the initial and 
final configurations are related by~\cite{blumental}:  
\begin{equation}
M_i(r_i) r_i = \left[M_b(r_f)+ M_{CDM}(r_f)\right] r_f\;.
\label{eq:adia}
\end{equation}
Here, $M_i(r)$, $M_b(r)$ and $M_{CDM}(r)$ refer, respectively, to the mass 
profile of the halo before the baryon infall (i.e. the form we assumed we can
infer from N-body simulation results), the baryon component as observed in 
the Galaxy today, and the cold dark matter component in its nowadays configuration.

In the opposite regime, a large transfer of angular momentum between the 
luminous and the dark components is assumed during the baryon infall. 
Mechanisms that induce a heating up of the population of cold dark matter 
particles sitting at the center of halos have been invoked to smooth out the 
inner CDM cusps, for which there is no evidence in real galaxies, or maybe 
even incompatibility, especially in the case of small dark matter-dominated 
objects, such as dwarf and low surface brightness galaxies~\cite{deblock}
(but see also ~\cite{vderb}).  As there is no mechanism intrinsic in CDM to remove 
central cusps, in several analysis the focus has been moved to the possible role of 
baryons. In a recent and, to some extent, extreme model~\cite{elzant}  baryons 
sink in the central part of halos after getting clumped into dense gas clouds:
the halo density profile in the final configuration is found to be described 
by a profile with a large core radius of the type proposed by Burkert~\cite{burkert}:
\begin{equation}
  \rho_{B}(r) = \frac{\rho_B^0}{(1+r/a)\,(1+(r/a)^2)}\;,
\label{eq:burk}
\end{equation}
a profile that has been shown to provide accurate fits of the rotation curves for 
a large sample of spiral galaxies~\cite{salucci}. 

The procedure we implement to select a halo model goes into several steps,
which we summarize briefly here and are described in detail 
elsewhere~\cite{halomod}: To a given halo, labeled
by a pair ($M_{vir}$,$c_{vir}$), with correlation between $M_{vir}$ and $c_{vir}$ 
of the form suggested by results in numerical simulations, we apply one of the two 
procedures for baryon infall sketched above. The gas and stellar terms in the
Galaxy are modeled as a superposition of components, i.e. a stellar bulge, bar 
and disk, plus two gaseous disks, whose morphologies are assumed in agreement 
with photometric studies in the Milky Way, and in analogy to the form observed in 
external galaxies. These components and the dark halo in its final configuration
are required to match available dynamical information: among others, we apply 
constraints following from the motion of stars in the Sun's neighbourhood, total mass 
estimates following from the motion of the outer satellites, and we require 
consistency with the Milky Way rotation curve and with measures of the optical 
depth towards the galactic bulge. 

As a result we extract two sample dark matter halo models:
i) a Burkert halo profile, with $M_{vir} = 1.3 \times 10^{12}\msun$ 
and $c_{vir} = 16$ (we use for the definition of $c_{vir}$ the convention
of Bullock et al.~\cite{bullock} with a slightly different normalization 
to the mean virial overdensity with respect to, e.g., Ref.~\cite{NFW}), 
and described by 
Eq.~(\ref{eq:burk}) with $a= 11.7$~kpc and a local halo density 
$\rho_B(r_0)= 0.34$~GeV~cm$^{-3}$;
ii) an adiabatically contracted profile, derived starting from a profile with the
non-singular form extrapolated in Ref.~\cite{n03} from simulations with 
the highest resolution so far,  and $M_{vir} = 1.8 \times 10^{12}\msun$ and 
$c_{vir} = 12$ (we label 
the final halo as N03 profile). The form of the latter is obtained numerically 
with Eq.~(\ref{eq:adia}) and we cannot give its explicit form: the local halo density
in this case is $\rho_{N03}(r_0)= 0.38$~GeV~cm$^{-3}$
and the profile has a pronounced cusp
towards the galactic center, driven in its most inner portion by the $r^{-1.8}$
singular scaling in the bulge profile~\cite{zhao}, which dominates over the
non-baryonic dark component. Adiabatic contraction makes the profile even 
steeper than the $r^{-1}$ cusp in the profile originally proposed by Navarro, 
Frenk and White~\cite{NFW}, and actually the result obtained starting from
this or shallower profiles is equivalent. The approximation of adiabatic 
contraction is assumed to be valid up to the radius of 1~pc, 
which corresponds to the range within
which the mass budget gets dominated by the central black hole~\cite{bhobservation}.
We consider a scenario for the black hole formation in which the CDM system
is perturbed and a core in the density profile of about 1~pc appears (see, e.g.,
~\cite{ulliobh,milo}), the opposite regime compared to the case of adiabatic growth 
of the black hole which induces the formation of a very dense spike within the
innermost parsec~\cite{gondolosilkbh}: the latter is essentially equivalent to 
adding a point dark matter source at the Galactic center, and can be treated 
separately.

The final step is to derive a velocity distribution function consistent with the 
dark matter density profile considered. For isotropic velocity distributions and in
the limit of spherically symmetric systems, the cases we restrain to in this 
analysis, the distribution function $F$ is determined univocally by the density 
profile through Eddington's formula~\cite{eddington}:
\begin{equation}
  F(\epsilon)=\frac{1}{\sqrt{8}\pi^2} \left[\int_0^{\epsilon} 
  \frac{d^2\rho_f}{d\Psi^2} \frac{d\Psi}{\sqrt{\epsilon-\Psi}} 
  +\frac{1}{\sqrt{\epsilon}}\left(\frac{d\rho}{d\Psi}\right)_{\Psi=0} \right]
\label{eq:eddi}
\end{equation}
where $\Psi(r)=-\Phi(r)+\Phi(r=\infty)$, with $\Phi(r)$ the potential for the 
Galaxy (i.e. including all components as determined from the dynamical 
information, in a toy model where the density profiles of stars and gas are assumed 
to be spherical as well, an approximation which may seem somehow drastic 
but has little influence on the final result), 
$\epsilon = - E +\Phi(r=\infty)=-E_{kin}+\Psi(r)$, and $E$ and $E_{kin}$, 
respectively, the total and kinetic 
energy. The functions $F(\epsilon)$ for our two sample models are computed 
numerically, after changing variable of integration in Eq.~(\ref{eq:eddi}) from 
$\Phi$ to the radius, $r$, of the spherical system.

Details of the two sample models we consider here are give in 
Ref.~\cite{halomod} and tabulated forms of these density profile and distribution
functions are also included in the latest release of the \ds\ package~\cite{ds}.
We just underline here that, for the first time, dark matter detection
is treated in a scheme where dark matter halo profiles are directly compared
to available dynamical constraints and fully self-consistent density profiles and
local velocity distributions are considered, so that self-consistent comparisons
between direct and indirect detection techniques are possible.

\section{Detection methods}

We will consider the discrimination and detection prospects with most of the
techniques that have been proposed so far for WIMP dark matter searches.
We describe briefly here the main underlying assumptions, and how we 
compare, in each class of experiments, with current sensitivities and with 
future ultimate-goal experiments.

\subsection{Direct detection}

Concerning direct detection, we will present our results in terms of the neutralino-nucleon
spin independent scattering cross sections (more precisely, we will plot 
the neutralino-proton cross section, as the neutralino-neutron one is analogous). Spin 
dependent cross sections are not considered here, because the detection prospects are less
encouraging. The reader should keep in mind that in the $\sigma_{\chi-p}$ which
is provided, is encoded, besides the dependence on SUSY parameters, a 
slight dependence on the underlying effective Lagrangian approach and on
nucleonic matrix elements. Regarding the first issue we use approximate heavy 
squark expansions rather than full one-loop gluonic interactions. On the latter, we take
a standard set of parameters~\cite{Gasser,SMC} (e.g., for the s-quark operator, 
the most important in determining the scattering cross section, we set the
parameter $f^p_{Ts} = 0.14$, slightly smaller than the value implemented
in other analyses, see~\cite{ds,paololars}
for details). The scattering cross sections are compared to the sensitivity level
reached by current direct detection experiments and to the projected final reach
of future large scale (one-ton-size) detectors (unfortunately, in the scheme we consider
we do not find any model which produces cross sections at the level needed to
provide an annual modulation effect compatible with the effect found by the 
DAMA Collaboration~\cite{dama}). For definiteness, we will refer to the 
EDELWEISS  2002~\cite{edelweiss} exclusion limit, and to the sensitivity
foreseen in the proposed XENON detector~\cite{xenon}. 
Both of these are recalculated for the
local velocity distribution functions in the two halo models considered here,
taking into account relevant effects, such as target materials and form 
factors, thresholds, exposures and backgrounds (sensitivity
curves for the standard Maxwell-Boltzmann velocity distribution have been derived
as well, and cross checked against published results, with which they agree).

\subsection{Neutrino flux from the Sun and the Earth}

The neutrino flux induced by pair annihilations of neutralinos trapped at the
center of the Sun, is computed according to the standard procedure described
in Ref.~\cite{joakimnt}, except for the treatment of the neutralino capture rate,
which is derived numerically implementing the expressions in \cite{gould-diff} 
which are valid for a generic velocity distribution function
(rather than the approximate expression
appropriate for a Maxwell-Boltzmann velocity distribution~\cite{jkg}, 
which are used by most authors), applied to the two cases considered here.
This allows for a fully consistent comparison between this signal and direct 
detection. The results are presented in terms of muon fluxes, above the threshold 
of 1~GeV, 
generated by the muon neutrino flux, and compared for reference with
the best exclusion curve among those provided by currently operating neutrino
telescopes. As the current best limits are from the SUPER-KAMIOKANDE Collaboration
in 2002~\cite{superKlimit} we will use these for the current sensitivity, and for the next generation of neutrino telescopes we will use the projected sensitivity of the km$^2$-size detector which is being built
by the ICECUBE Collaboration~\cite{icecube}. For these projected sensitivities, we will use the expected sensitivity for the particular annihilation channel that dominates the neutralino annihilation, i.e.\ the sensitivities will be worse for soft channels (e.g.\ $b \bar{b}$) than for hard ones (e.g.\ $W^+W^-$, $Z^0 Z^0$ and $t \bar{t}$).
The neutrino flux induced by neutralinos captured by the Earth has been computed as 
well. However, our results show that the expected signals are always orders 
of magnitude below the expected sensitivities. One should also remember that 
the Earth signal is highly correlated to the size of the spin-independent 
coupling to nucleons (in the same way the Sun signal is correlated to
the spin-dependent and spin-independent coupling), and we find that direct detection seems always
more promising. We are not going to consider this signal further.

\subsection{Antimatter fluxes}

Neutralino pair annihilations in the Galactic halo may produce a significant
amount of antimatter, even at the level of the standard secondary flux 
generated in the interaction of primary cosmic rays with the interstellar medium.
We consider here neutralino induced fluxes of antiprotons, positrons and 
antideuterons. The prediction for these fluxes consists of several steps: our particle
physics setup fixes the pair annihilation cross section at zero temperature
$\sigma_{\rm ann}v$ (i.e. the probability, at the present time, for non-relativistic neutralinos to annihilate), and the branching ratios for the various annihilation channels. 
The fragmentation and/or decay of these annihilation products give rise to the stable 
antimatter species. We have modeled this process with the \code{Pythia} \cite{pythia} 
6.154 Monte Carlo code, 
simulating for each allowed two-body final state and for a set of 18 neutralino 
masses, except for $\bar{D}$ sources for which we have implemented the prescription 
suggested in Ref.~\cite{dbar} to convert from the $\bar{p}$-$\bar{n}$ yields. 
The strength of local sources scales with the number density of neutralino 
pairs locally in space, i.e., in terms of the dark matter density profile,
with $1/2\,(\rho_{\chi}(\vec{x}\,) / {m_{\chi}})^2$, where $m_{\chi}$ is the neutralino 
mass. The choice of halo profile enters then critically: large enhancements in the
source functions are provided by the adiabatically contracted profile which is cuspy
towards the Galactic center.

The next step is to model the propagation of charged cosmic rays through the 
Galactic magnetic fields. We consider a two-dimensional diffusion model, in 
which reacceleration effects are not included explicitly, but mimicked through 
a diffusion coefficient which takes the form of a broken power law in rigidity, $R$, 
\begin{eqnarray}
D = D_0\,(R/R_0)^{0.6}\;\;\;\;\;\;\;\;& {\rm if}& R \ge R_0 \nonumber \\
D = D_0\;\;\;\;\;\;\;\;\;\;\;\;\;\;\; & {\rm if}& R < R_0\,.
\label{eq:diff}
\end{eqnarray}
E.g., in Ref.~\cite{strmosk} such a form has been tested with the 
\code{Galprop}~\cite{galprop} propagation code, showing that, with an appropriate
choice of parameters, it is possible to reproduce fairly well the ratios of primary 
to secondary cosmic ray nuclei. We adopt here the same parameter setup, namely
Eq.~(\ref{eq:diff}) with $D_0 = 2.5 \times 10^{28}$~cm$^2$~s$^{-1}$ and $R_0=4$~GV, 
in a cylindrical diffusion region of radius equal to 30~kpc and half height equal to
4~kpc, plus a galactic wind term. In the case of antiprotons and antideuterons this is
interfaced into the diffusive-convective code illustrated in Ref.~\cite{pbarpaper}, 
which neglects energy losses (the particle is removed whenever a scattering with a nucleus takes place). For positrons, we instead use the code developed in
Ref.~\cite{epluspaper}, improved and extended to allow for the implementation
of a diffusion coefficient in the form of Eq.~(\ref{eq:diff}), and to keep a full 
two-dimensional structure. In this code there is no convective term but there
is a term accounting for positron energy losses due to inverse Compton scattering
on starlight and the cosmic microwave background.

The quality of the data on the local antiproton and positron cosmic ray flux has kept
improving in recent years. We will compare the predicted fluxes to the antiproton data 
collected by the BESS experiment during its flights in 1997, 1998, 1999 and 2000
\cite{bess} with fairly good statistics in the energy range between 180 MeV and 
4.2 GeV, and by the CAPRICE experiment during its 1998 flight~\cite{capricepbar} 
in the range between 3 and 50 GeV\@.  For the positron fluxes, we refer to data published
by the  HEAT Collaboration on their 1994-1995 flight~\cite{heat}, by the CAPRICE team 
regarding a flight in 1994~\cite{capriceeplus}, plus data obtained by MASS-91 in 
1991: the overall energy range covered by these measurements extends from
460 MeV to 34.5 GeV. We have chosen not to include in our analysis data which
have been reported just as antiproton or positron fractions (rather than absolute 
fluxes) and datasets such as the one on positrons from the AMS test 
flight \cite{amseplus} mapping a low energy interval in which a primary 
neutralino-induced contribution is not likely to be present. 

To compare with these 
data it is necessary to include the effect of solar modulation, i.e. the effect of
propagating fluxes from interstellar space to our location inside the solar  system and against the solar wind.  To sketch this effect, we implement the 
one parameter model based on the analytical force-field approximation by Gleeson \& 
Axford \cite{GleesonAxford} for a spherically symmetric model. The solar modulation
parameter, sometimes dubbed Fisk parameter $\Phi_F$~\cite{fisk}, is for simplicity
assumed to be charge-sign independent, and is derived fitting the measured
local proton cosmic-ray flux to the solar activity corresponding to the period
of data taking of each of the considered datasets.
Finally, it is also necessary to consider a model for the secondary antimatter 
fluxes, which play the role of backgrounds in our analysis. For this task we refer
again to the  \code{Galprop}~\cite{galprop} code, running the code in the same 
setup we use for the signals and for the current best estimate for proton cosmic-ray
flux, and extracting the output on secondary antiprotons and positrons. For
both species, these backgrounds provide excellent fits of the data: we obtain,
for background only, a reduced $\chi^2$ equal to 0.82 for antiprotons and to 0.95
for positrons.

Future measurements of antimatter in space, such as with the PAMELA satellite
detector~\cite{pamela} and the AMS spectrometer~\cite{ams} on board the 
International Space Station Alpha (ISSA), with instruments that will operate over 
very long  exposures (tentatively, at least 3 years), will provide data on an energy 
range which is wider than that of present data and with much better statistics.
To address the perspectives of discrimination of the signal against the background, 
we follow the approach outlined in Ref.~\cite{stefanopiero} and define the quantity:
\begin{equation}
I_{\Phi} \equiv \int_{E_{min}}^{E_{max}} dE \, \frac{\left[\Phi_s(E)\right]^2}{\Phi_b(E)}\,,
\end{equation}
where $\Phi_s(E)$ and $\Phi_b(E)$ refer, respectively, to the signal and the background flux, and the integral extends over the energy interval in which the 
integrand is non negligible. This represents the continuum limit, up to an overall
factor which accounts for the  exposure times effective area of a given future
experiment, of the $\chi^2$ variable of the form:
\begin{equation}
\chi^2=\sum_{i} \frac{\left(N_s^i+N_b^i-N_o^i\right)^2}{\left(\Delta N_o^i\right)^2}
\label{iphi}
\end{equation}
under the assumption that, in each energy bin $i$, the number of signal events $N_s^i$
is subdominant with respect to the number of background events $N_b^i$, which in turn nearly matches the number of observed events $N_o^i$, and that errors can be 
approximated as statistical errors, i.e. 
$\Delta N_o^i \simeq \sqrt{N_o^i} \simeq \sqrt{N_b^i}$. We have assumed as well
that the measurement covers the energy range where the bulk of the signal is 
expected, and that the background is known, as one can (optimistically) expect
from future determinations of the propagation parameters from, say,  high
precision measurements of ratios of secondaries to primaries for several light 
nuclei. With these hypotheses, we will indicate up to about what level of the 
parameter $I_\Phi$, the PAMELA detector will be able to reject the presence of a 
signal contribution in  the antiproton and positron flux.

The case for antideuterons is different because in this case, restricting to
a low energy window, the background flux is expected to be negligible~\cite{dbar},
and even detection of 1 event would imply discovery of an exotic component. 
Regarding the detection prospects, we will consider,
as the ultimate reach for an experiment in the future, that of the gaseous antiparticle 
spectrometer (GAPS) \cite{GAPSproposal}. This is a proposal for an instrument
looking for antideuterons in the energy interval 0.1-0.4 GeV per nucleon, with
estimated sensitivity level of $2.6\times10^{-9}\textrm{m}^{-2}\textrm{sr}^{-1}\textrm{GeV}^{-1}\textrm{s}^{-1}$, to be placed either on a satellite orbiting
around the earth or on a probe to be sent into deep space. 

\subsection{Gamma rays}

A certain amount of gamma-rays is also produced as a result of the fragmentation
and decay of annihilation products from neutralino annihilations in the halo (at the same time
a monochromatic component may be present, due to two-body annihilation states containing
a photon; this component is not considered here, see, e.g., \cite{BUB}, for a discussion
of detection prospects). The induced flux is generally small, unless one deals with a dark matter halo profile with sharp density enhancements locally in space. As one of the two profiles we are considering here is cuspy towards the Galactic center (GC), we will compute the neutralino-induced $\gamma$-ray flux in that direction. 

The EGRET experiment, on the Compton Gamma-Ray Observatory has resolved a $\gamma$-ray source towards the GC~\cite{MH}, tentatively extended ($\sim 1.5^{\circ}$, of the order of the EGRET angular resolution) rather than point-like, and with a spectrum that is sensibly harder than the spectrum expected for the diffuse $\gamma$-ray flux due to  the interaction of primary cosmic rays with the interstellar medium. Quite intriguingly, the measured flux shows the distortion that the diffuse flux would have in case of a WIMP-induced component~\cite{cesarini}. We will compare the fluxes we find in our framework with the intensity and spectrum of the EGRET $\gamma$-ray source. Note, however, that alternative tentative explanations for this source have been proposed~\cite{pohl}, and that even the identification of the location of the source with the GC has been questioned~\cite{hooper}. Should one of these two issues be confirmed by the upcoming observations with the GLAST satellite \cite{glast}, more stringent limits than those shown here could be derived.

More recently, observations at higher energies of a $\gamma$-ray flux in the GC direction with ground-based Air Cerenkov Telescopes~\cite{cangaroo,whipple} have been reported. Given the neutralino mass range we will consider here, we find that the EGRET measurement gives always a more stringent bound.

\section{Results and discussion}

In line with most previous analyses, we will sample the 5-dimensional mSUGRA
parameter space choosing a few values of $\tan \beta$ and $A_0$, and varying
$m_{1/2}$ and $m_0$ for both $sign(\mu)$.  We consider three different
regimes: i) $m_0 \lsim m_{1/2}$ and  $A_0=0$; ii)  small $m_{1/2}$, moderate
$m_0$ and large $A_0$; and iii) $m_0 \gg m_{1/2}$. For each of these regimes we
find the isolevel curves for $\Omega_{\chi} h^2 = 0.103$, and select the parts of these curves that fulfill the accelerator constraints.

\subsection{Bulk, slepton coannihilation and funnel regions}

In the regime at $m_0 \lsim m_{1/2}$ and  $A_0=0$ 
there are three distinct frameworks which drive the relic density of binos to be in the
cosmologically favored range. Starting at small $m_{1/2}$ and small $m_0$,
we find the ``bulk'' region, where the relic density of pure binos is set by their 
annihilation strength into fermions mediated by a sfermion; it is a region, however, 
which is almost entirely excluded by accelerator constraints, see, e.g., the 
discussion in Ref.~\cite{baer}. Going to larger $m_{1/2}$ there are two 
possibilities: for small or moderate values of $\tan\beta$ the curves with a fixed
value of the relic abundance enter into the slepton coannihilation tail, a thin strip 
on the border with the region where a stau is the LSP and the relic abundance 
gets dominated by coannihilation effects with the lightest stau, as well as with the 
lightest selectron and smuon (this region was recognized in Ref.~\cite{Ellisstau1} 
and the relic density calculation has been discussed in several recent
analyses, including~\cite{Ellisstau2,GLP,roszkowski,bbb,bbb7.64,coann}). 
As sample cases of this regime, we have searched for relic density isolevel curves 
for $\tan\beta=10$ and $\tan\beta=30$. As already mentioned, for large $\tan\beta$,
one hits instead the region where the pair annihilation of neutralinos goes on the
s-channel resonance of $H_0^3$ and/or $H_0^1$: $\Omega_\chi h^2$ isolevel curves
get two separate branches, at slightly larger and slightly smaller values of $m_0$
with respect to the value that, at a fixed $m_{1/2}$, drives the neutralino mass to be
one half of the Higgs mass; the two branches merge at the largest allowed values of  $m_{1/2}$ (and hence of the neutralino mass) for a given value of the relic density. This is more evident for
$sign(\mu)=-1$, as in this case a sharp resonance appears in the neutralino
LSP region even for values of $\tan\beta$ around 40 or so, 
while for positive $\mu$ one has to go
to very large $\tan\beta$. We consider the case $\tan\beta=53$, $sign(\mu)=+1$    
and $\tan\beta=46$, $sign(\mu)=-1$; for both of them the isolevel curves for $\Omega_\chi h^2$
in the cosmologically interesting band, start, at small $m_{1/2}$, in the bulk region, 
going to larger $m_{1/2}$, enter the upper resonance branch, and then bend down 
into the resonance branch with lower $m_0$, going now to smaller $m_{1/2}$, and 
finally falling into the slepton coannihilation tail, as for the case with 
lower $\tan\beta$.
Although we have derived, for all $\tan\beta$ and  both $sign(\mu)$, the isolevel 
curves in the full plane, we will have to take into account that the \bsg\ constraints
will cut a large portion of the parameter space, especially at large $\tan\beta$ and
negative $\mu$. Also in the case of $sign(\mu)=-1$ the SUSY contribution to $(g-2)_\mu$
is negative. It is still within our broad range of allowed values, but slightly disfavoured.

\begin{figure}[t]
\centerline{
\includegraphics[width=0.49\textwidth]{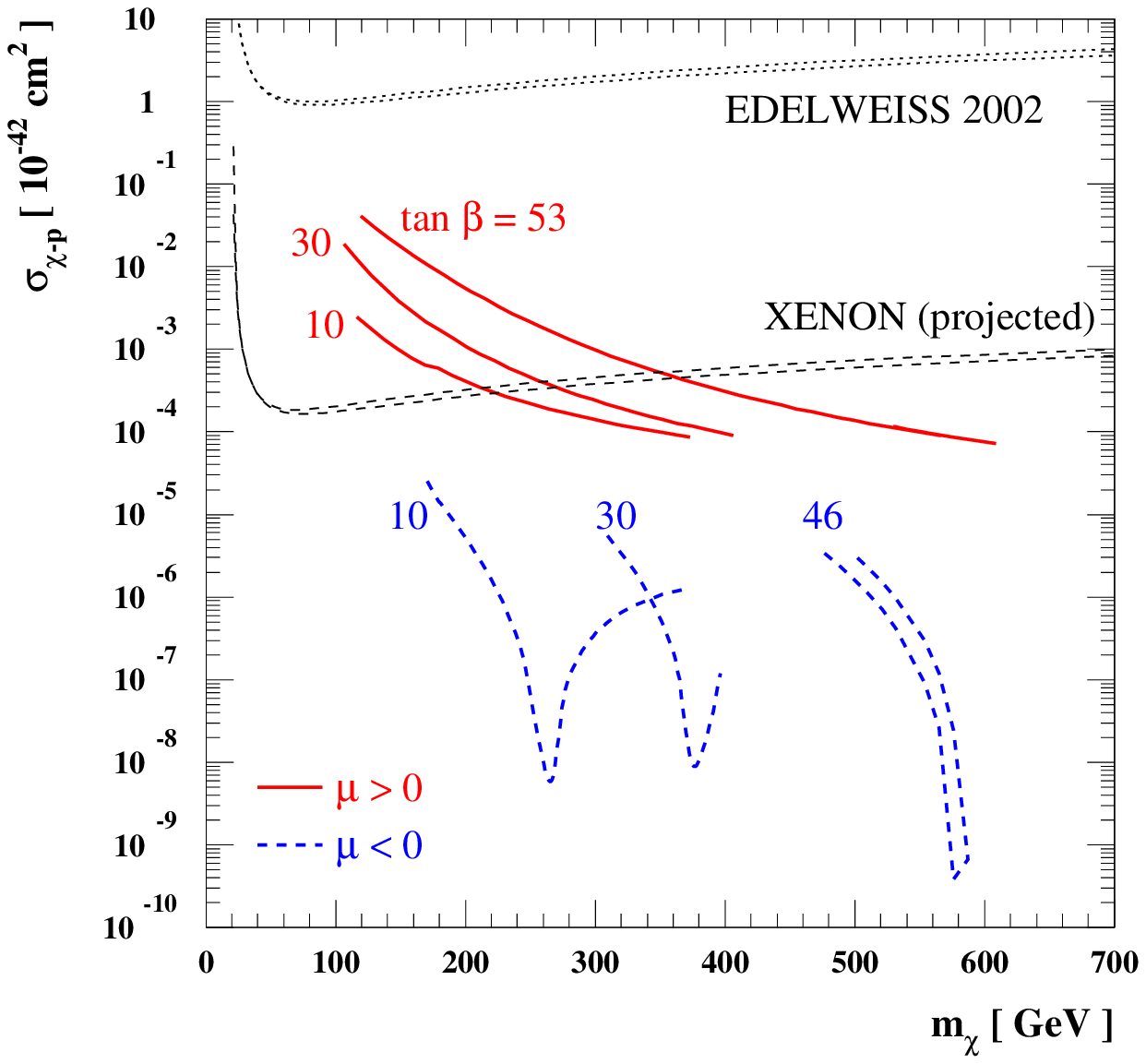}
\includegraphics[width=0.49\textwidth]{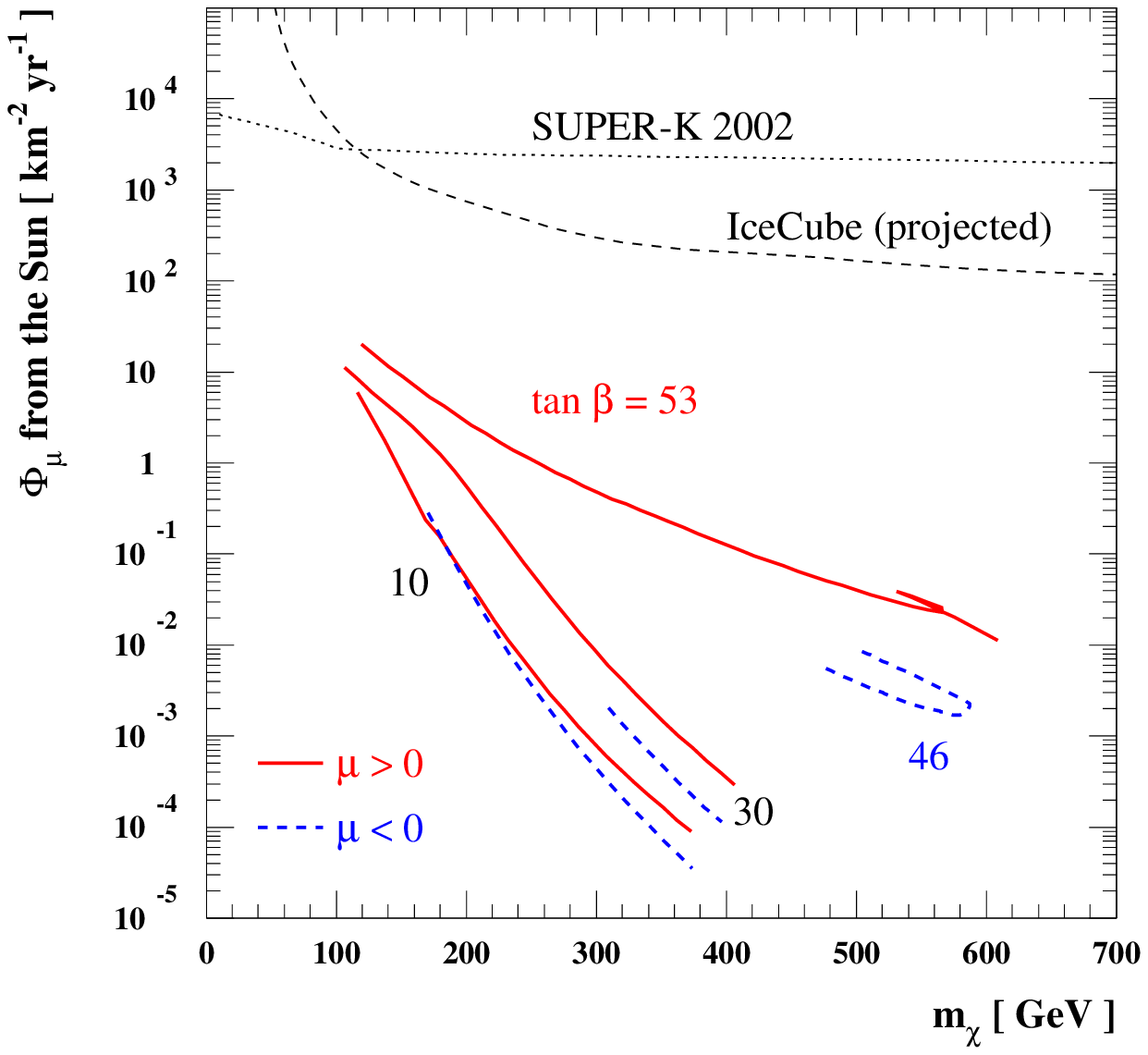}}
\caption{The expected direct detection spin independent scattering cross section (left panel) and neutrino-induced muon fluxes from the Sun (right panel) for a few sample 
cases at $m_0 \lsim m_{1/2}$. Along each of the curves we have fixed $\Omega_\chi h^2 = 0.103$. The lower curves for the direct detection limits are for the N03 profile, and the upper for the Burkert profile. The predicted muon fluxes are for the Burkert profile, with the N03 fluxes being only marginally larger.}
\label{fig1}
\end{figure}

Sitting on the isolevel curves for $\Omega_\chi h^2 = 0.103$, we show our results in 
figures where we plot the neutralino mass on the horizontal axis, in the range allowed
in the lower end by the accelerator constraints, the $H^0_2$ mass limit for positive
$\mu$ and  \bsg\ for negative $\mu$, and in the upper end by the fact that the
isolevel curve hits the region where the stau is the LSP.
The neutralino mass labels univocally the location
on an isolevel curve, except along a portion of the funnel
stripe (as mentioned above). In Fig.~\ref{fig1} we show the expected spin independent scattering cross 
section and muon rates induced by the neutrino fluxes from the Sun. Regarding
the first, we notice that as the squarks are heavy, the main contributions to the
cross section always come from diagrams with t-channel exchange of 
$H^0_2$ and $H^0_1$ (which are almost decoupled from each other, and then, in their turn, coupled
mostly, respectively, to u-type SM fermions and d-type SM fermions). The size of these 
two contributions are comparable. The one through $H^0_1$ rises with $\tan\beta$, as the $H^0_1\,d\,\bar{d}$ vertex scales like $\sqrt{1+(\tan\beta)^2}$. The two contributions have  a constructive interference for positive $\mu$, while they have a destructive interference for negative $\mu$, giving very small values of the cross section. Even for positive $\mu$
and large $\tan\beta$, the scattering cross sections are rather small, at least a factor of 50 below
the current sensitivities; this feature is due to
the fact that we are dealing with rather pure bino-like neutralinos, while the 
neutralino coupling to the CP-even Higgs is roughly scaling with  $Z_g (1-Z_g)$,
with $Z_g$ being the gaugino fraction, i.e. $Z_g \simeq |N_{11}|$ in our case:
the coupling is suppressed for models with $Z_g$ too close to 1. Considering the detection
prospects for the future, for $\mu >0$ there is a fair range of models which lie above the
projected exclusion curve of the XENON experiment (the two dashed curves refer
to the two halo model considered, with the lower one being the one for the N03 
profile, mainly because the local halo density is larger in that case), but the high mass
end, for each $\tan\beta$, will not be tested. For negative $\mu$, none of the allowed models 
will be testable. Regarding the neutrino fluxes from the Sun, we have compared with the current limit from Super-Kamiokande \cite{superKlimit} (derived for an average type of neutrino energy spectrum). We have also compared with the expected sensitivity for IceCube \cite{icecube}, derived for a soft annihilation channel as the dominating annihilation channel is $b \bar{b}$ for these models. As Fig.~\ref{fig1}
shows, the rates are too low even for future large size neutrino telescopes. The flux
displayed is for the Burkert profile, but even for the N03 case the increase is marginal.
The fluxes are much lower than in more favorable scenarios, because the spin dependent cross section is lower than in these cases (and hence the capture rate is lower), as 
it is mediated, for pure binos, mainly by 
squark s-channel exchanges, and squarks in this scenario are rather heavy.

\begin{figure}[t]
\centerline{
\includegraphics[width=0.49\textwidth]{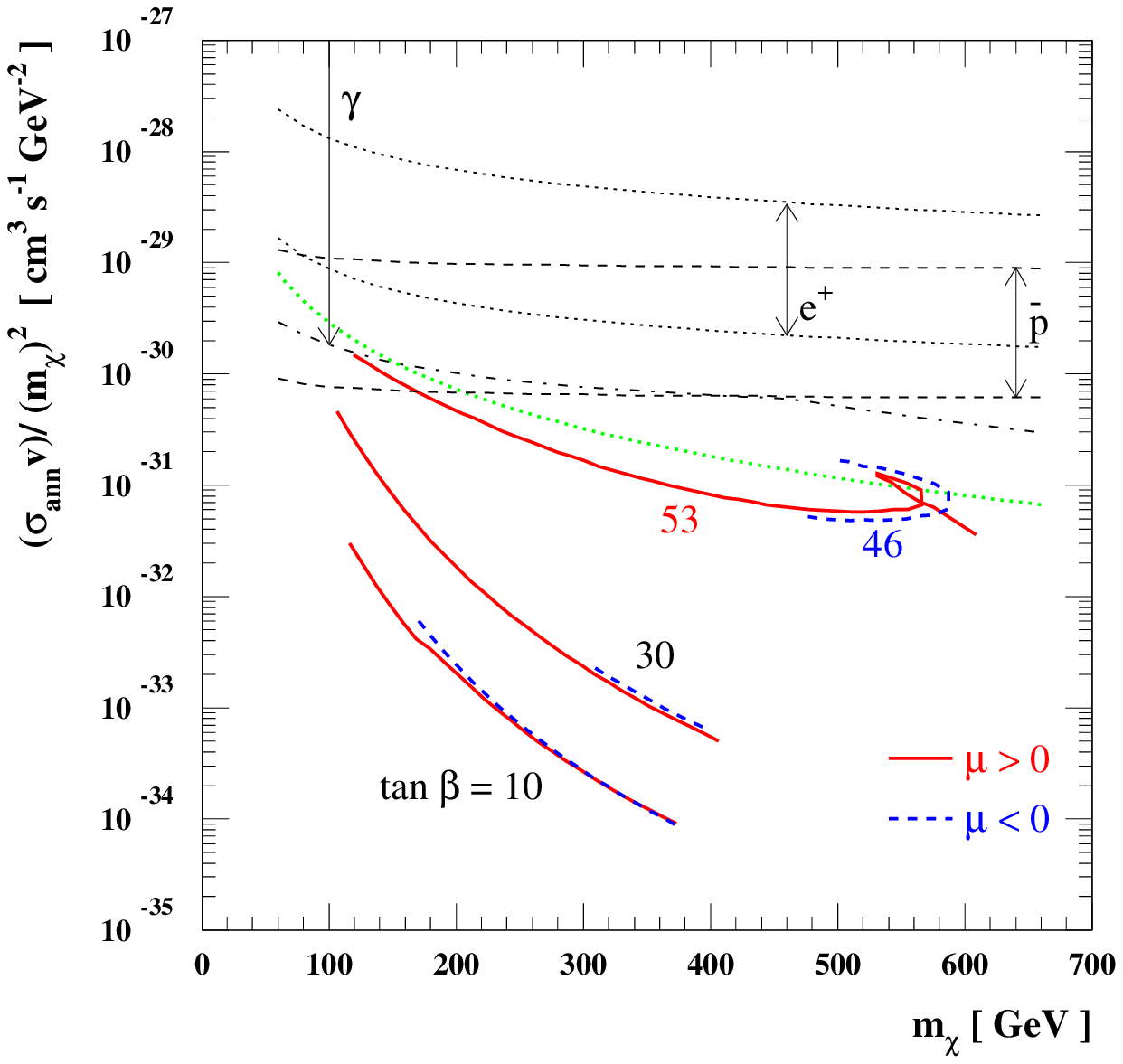}
\includegraphics[width=0.49\textwidth]{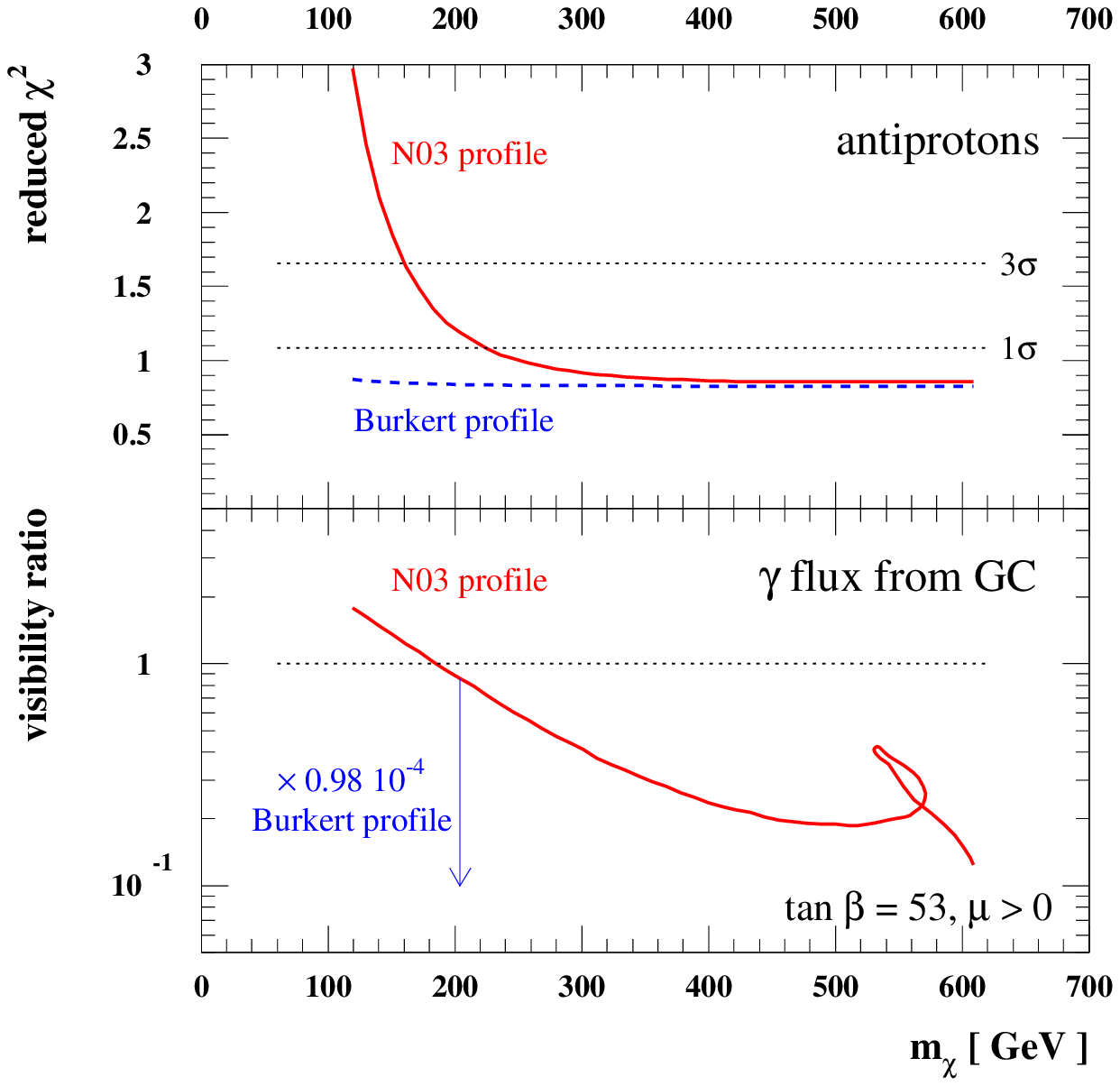}}
\caption{In the left panel we show the annihilation cross section divided by the neutralino mass squared for a few sample cases in the limit $m_0 \lsim m_{1/2}$. In the right panel we compare the predicted antiproton and gamma-ray fluxes against current data, see the text for details.
}
\label{fig2}
\end{figure}

The case for indirect signals from halo annihilation is introduced in Fig.~\ref{fig2}.
As a preliminary remark, we stress that we do not expect correlation of the halo rates
with spin independent or dependent scattering cross sections, or better, we may find
large rates for large cross sections (and actually, to some extent, we do) but it is never the same effect driving them to be large, i.e. crossing symmetry is never at work here.
In the left hand side of Fig.~\ref{fig2}, we plot the annihilation rate at zero temperature divided by the neutralino mass squared, which is the quantity appearing in the source function of all halo rates. Naively, implementing an approximate solution of the Boltzmann equation describing neutralino decoupling in the early Universe in which temperature dependences in the annihilation cross section are neglected, $\sigma_{ann}v$ scales
with the inverse of the relic abundance~\cite{jkg}:
\begin{equation}
\sigma_{ann} v \sim \langle \sigma_{ann} v \rangle \sim \frac{3 \times 10^{-27} {\rm cm}^3 
{\rm s}^{-1}}{\Omega_\chi h^2}\,.
\label{simpleomega}
\end{equation}
We are working at a fixed $\Omega_\chi h^2 = 0.103$; the corresponding curve is shown
as a green dotted line in the figure. If this approximation holds, the curves for the models we are considering should align on top of it, while we see that this is marginally true just for the
isolevel curve following a trajectory within resonances (which are very broad and hence thermal effects are not dramatic). For the other curves two effects come in: first there is the S-wave suppression of the cross section for the annihilation of non-relativistic Majorana fermions into light fermions, an effect that appears already for small neutralino masses; on top of this, going to larger masses, the thermally averaged annihilation cross section setting the relic abundance gets dominated by coannihilation effects. In this case, we can see that the scaling with mass of $(\sigma_{ann} v)/m_\chi^2$ on the curves referring to the $\tan\beta=10$ or 30 cases is much more rapid than $1/m_\chi^2$. In the figures are plotted also curves which give a feeling for the constraints from current data. They are calculated under the hypothesis that the $b\bar{b}$ annihilation channel dominates, a very good approximation, especially  for the funnel case. The dashed lines correspond to approximate $3 \sigma$ exclusion curves based on current antiproton data, for the two halo model considered, the lower curve corresponding to the N03 profile (and assuming the propagation model we picked is the correct one); they are nearly flat because we have factored out most of the dependence on mass in the signal in the quantity we plot on the vertical axis. The dotted lines are the analogous 
for positrons, and we see that, in this case, the curves are well above the predictions even for the N03 profile. Finally, the dash-dotted lines indicate cross sections for
which the EGRET excess from the GC is better fitted by a neutralino-induced component in case of the N03 profile plus a (normalization free) diffuse background component (see~\cite{cesarini} for details on the background and on how this fit is done); the case for the Burkert profile does not fit into the range of cross sections displayed on the vertical scale.
In the lower mass tail of the case with $\tan\beta = 53$ and the N03 profile, it looks that current data may put a constraint on the model. This is shown on the upper right-hand side of Fig.~\ref{fig2}, where we plot the reduced $\chi^2$ for the antiproton flux with respect to current data, and the ratio between the predicted gamma-ray flux in the energy bin between 4~GeV and 10~GeV and the flux measured by EGRET in the same bin (which is the highest one for the EGRET source, and turns out to be the one setting the tightest constraint). For the
N03 profile the lowest mass range seems to be excluded by both signals; note however,
that the $3 \sigma$ and $1 \sigma$ exclusion lines we plot for the antiprotons do not take into account uncertainties in the propagation models, and should then probably be slightly relaxed (but most likely not up to a reduced $\chi^2$ of 3), and that the estimate of the gamma-ray flux is very sensitive (much more than the antiproton flux) to details in the halo profile towards the GC: for the Burkert profile, the flux dies out to 4 orders of magnitude below the limit  (for reference, in the language of Ref.~\cite{BUB}, $\langle J \rangle$, the line of sight integral angularly averaged over a cone with a $1.5^\circ$ aperture,  is equal to  about $7 \times 10^4$ for the N03 profile considered here).

\begin{figure}[t]
\centerline{
\includegraphics[width=0.49\textwidth]{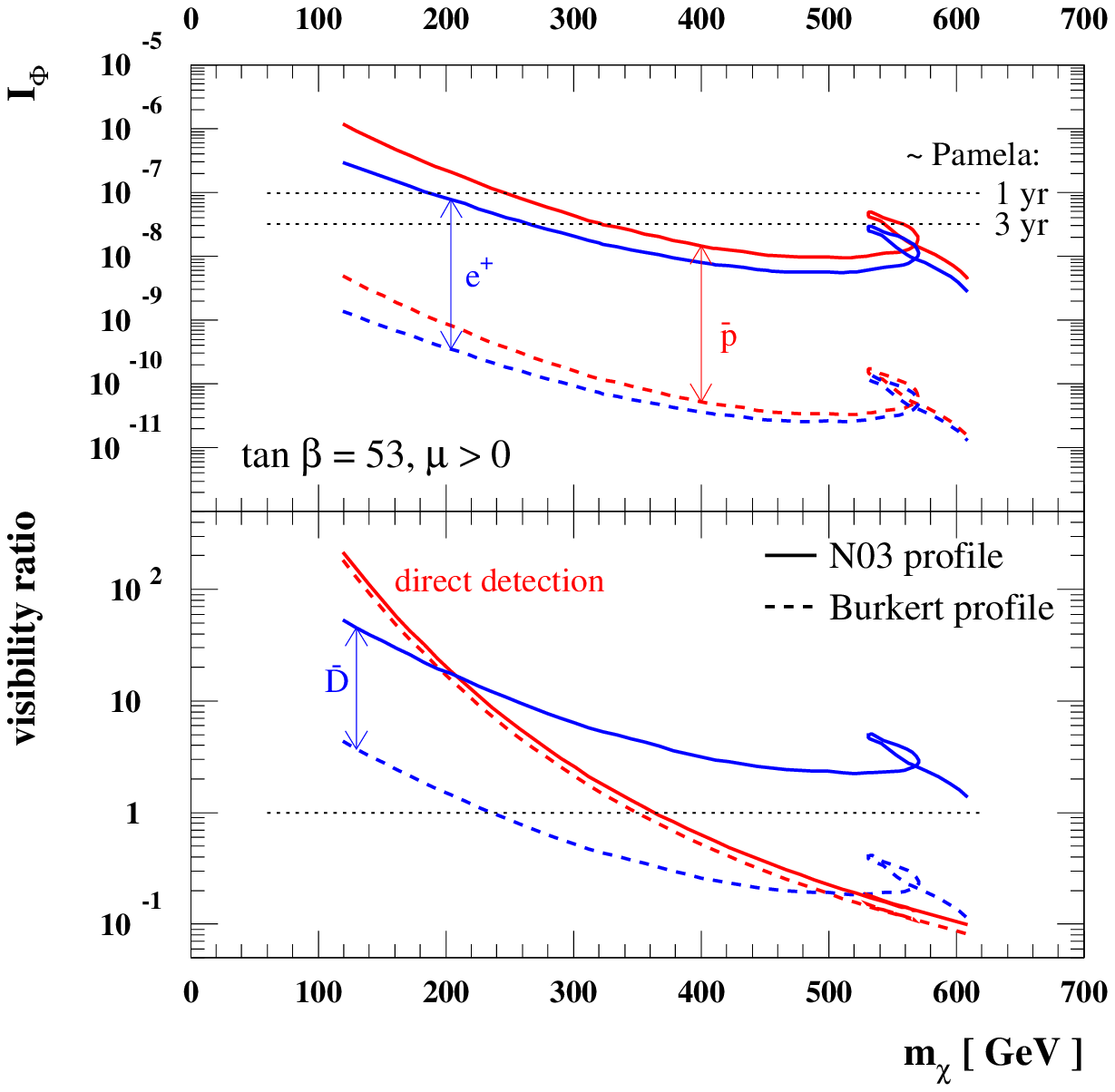}
\includegraphics[width=0.49\textwidth]{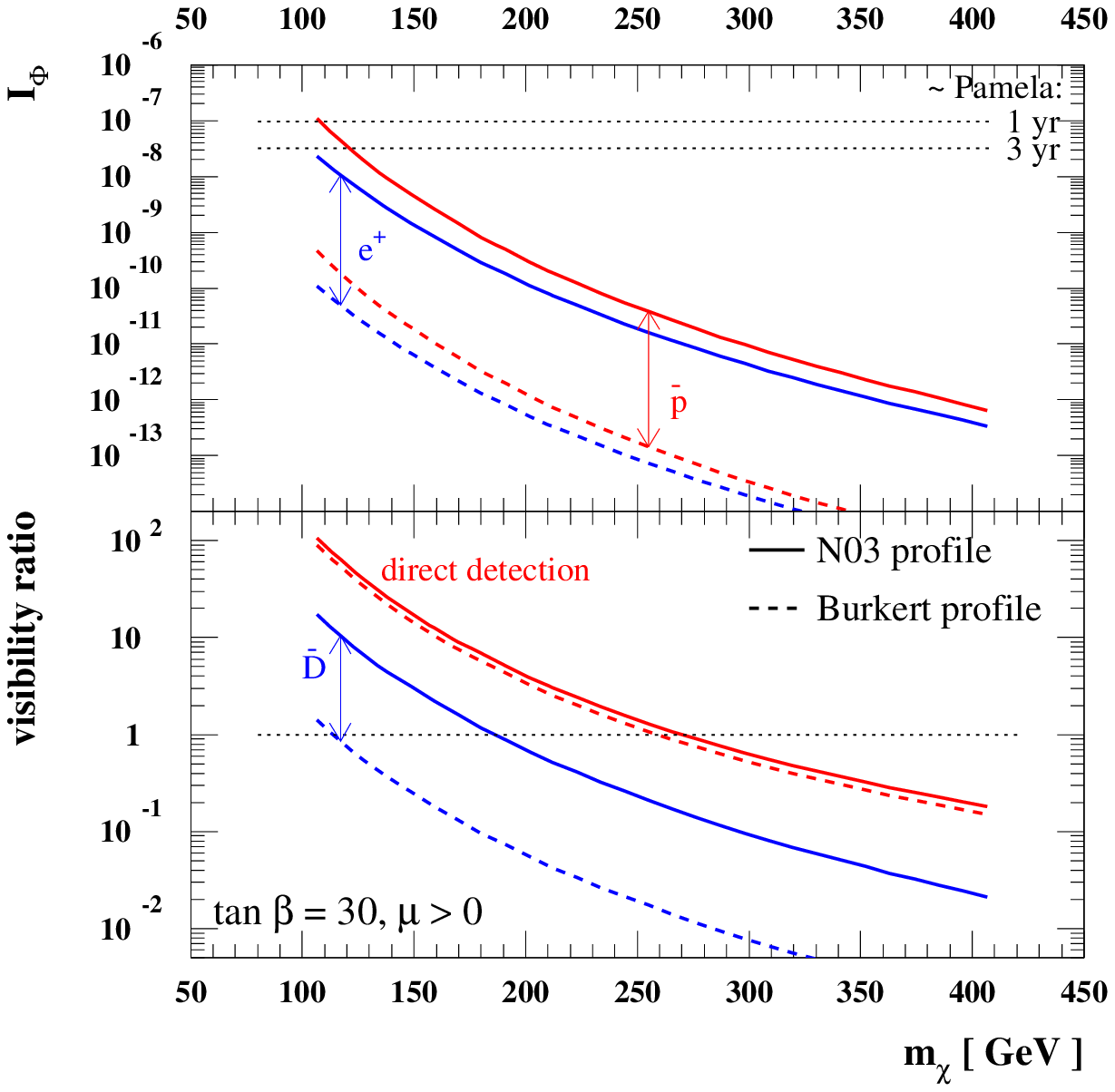}}
\caption{Perspectives to detect dark matter within the mSUGRA framework and the 
$m_0 \lsim m_{1/2}$ regime. Dotted curves in the upper part of the two figures
refer to approximate 3 $\sigma$ exclusion curves; in the lower parts of the figures, models with visibility ratios larger than 1 should be considered detectable in future experiments. Models with $\Omega_\chi h^2 = 0.103$ and two different values of $\tan\beta$ are displayed.}
\label{fig3}
\end{figure}

In Fig.~\ref{fig3}, we address the detection prospects for future experiments, for the
case $\tan\beta = 53$ and $\tan\beta = 30$, both with positive $\mu$. In the lower part of the figures are plotted the ratio between the spin independent scattering cross section and the projected sensitivity of the XENON proposed experiment, for the two halo models, as well as the predicted $\bar{D}$ flux, at the earth position and in a period towards solar maximum, over the sensitivity of GAPS (on a satellite close to Earth). As already mentioned, direct detection looks rather promising, and also very promising seems, for the N03 profile, but partially also for the Burkert profile, the measurement of the $\bar{D}$ flux (which would be even more promising should the experiment be performed with a detector placed on a probe sent into deep space). In the upper part of the figures, with the same mass scale, we plot the quantity $I_\Phi$ defined in Eq.~(\ref{iphi}), in units of cm$^{-2}$ s$^{-1}$ sr$^{-1}$,
compared with the 3 $\sigma$ exclusion curves in this quantity for an experiment with
an effective area of the size of PAMELA~\cite{pamelasens} (see~\cite{stefanopiero}
for details) and an exposure of 1~yr or 3~yr. As can be seen, these indirect channels looks less promising, and the positron channel seems always worse than
the antiproton one (but it should be kept in mind that we are assuming that the background is known; if a signal is indeed present, it might be easier to spot it in the positron 
flux at relatively high energy, rather than in the antiproton flux at low energy).

\subsection{Stop coannihilation region}

\begin{figure}
\centerline{\epsfig{file=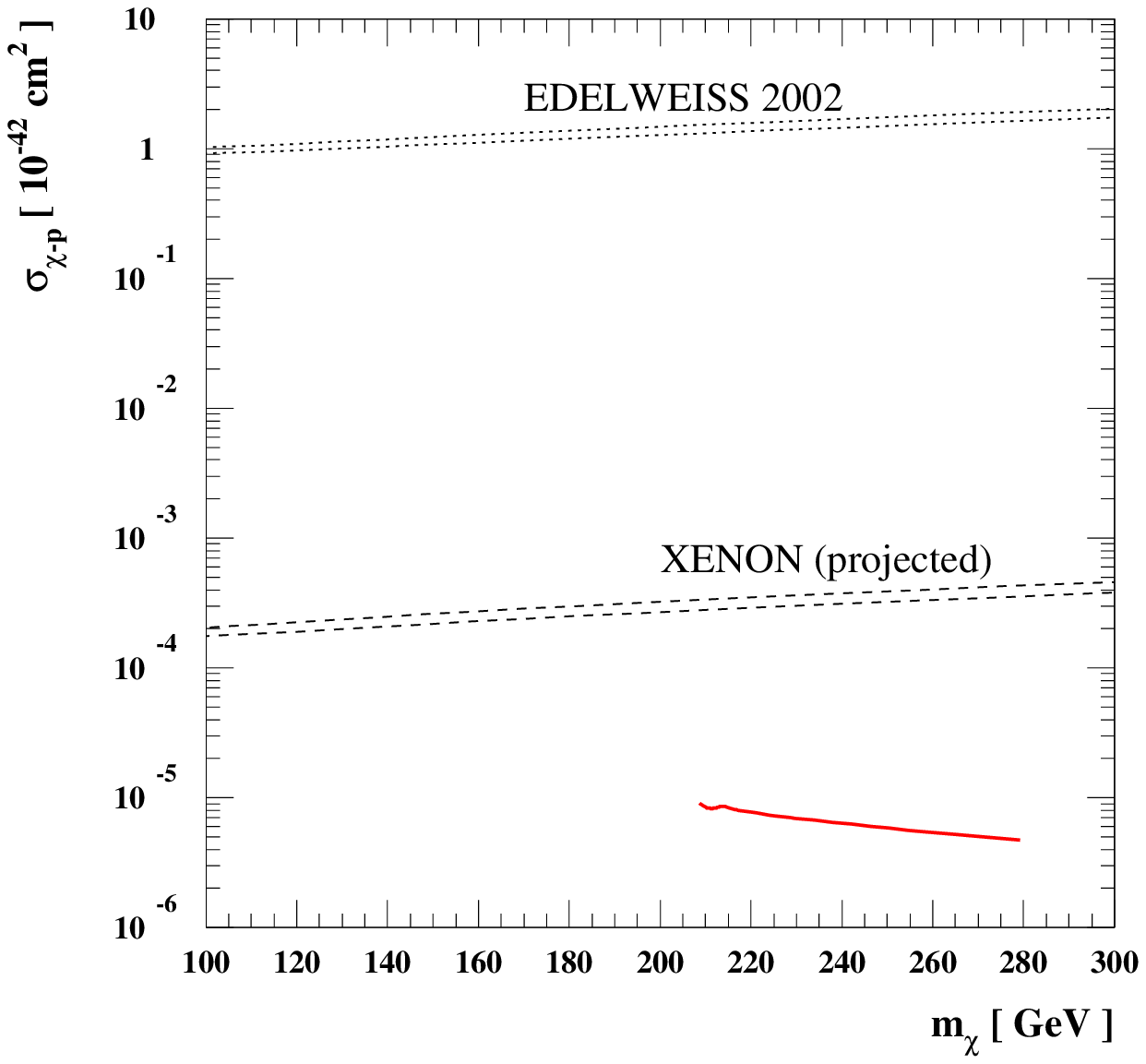,width=0.49\textwidth}
\epsfig{file=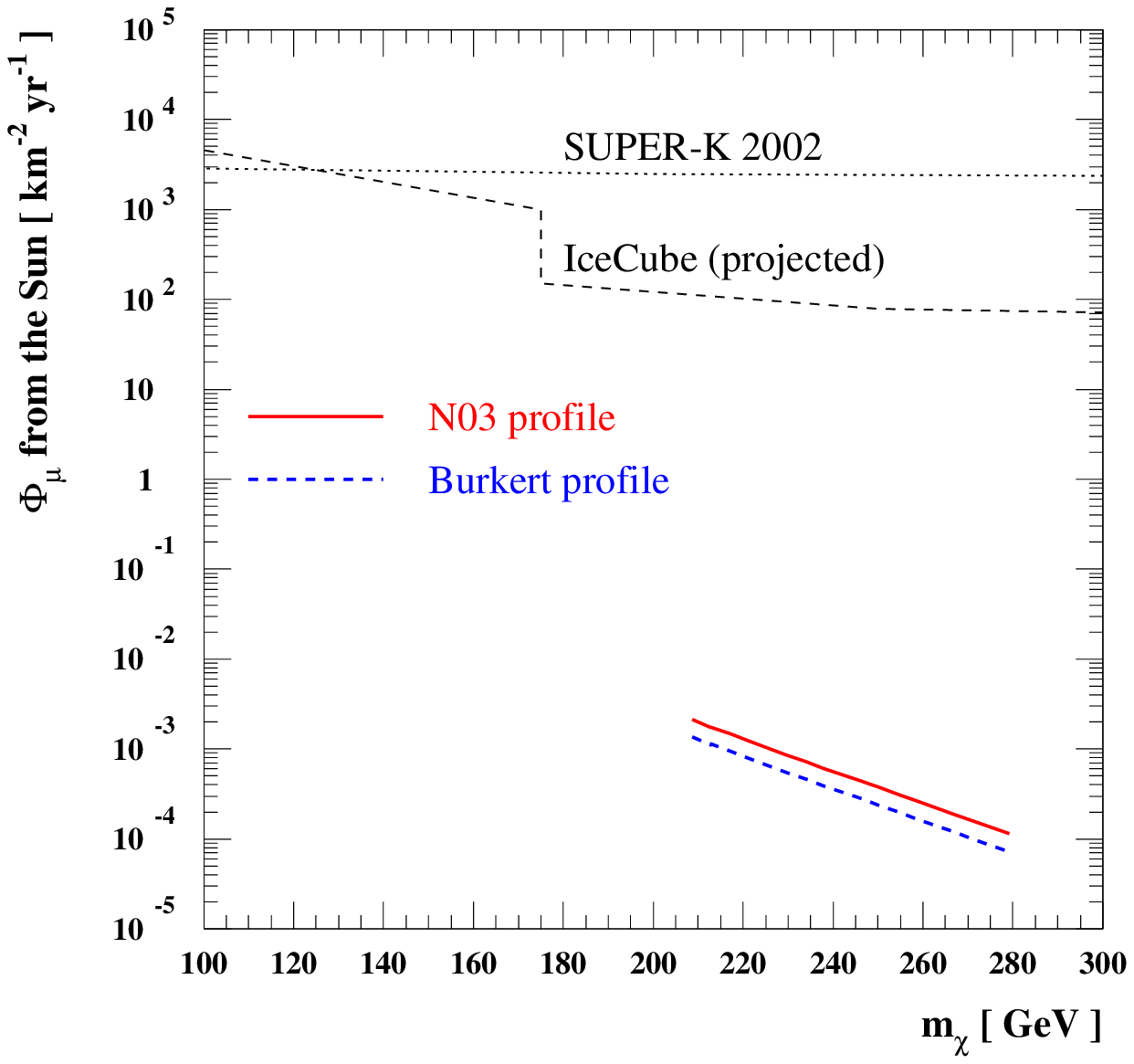,width=0.49\textwidth}}
\caption{The direct detection scattering cross section (left panel) and the
expected neutrino-induced muon fluxes from the Sun (right panel) for our chosen stop coannihilation region. The curves are analogous to those in Fig.~\protect\ref{fig1}.}
\label{fig:stopmuflux-sigma}
\end{figure}

\begin{figure}
\centerline{\epsfig{file=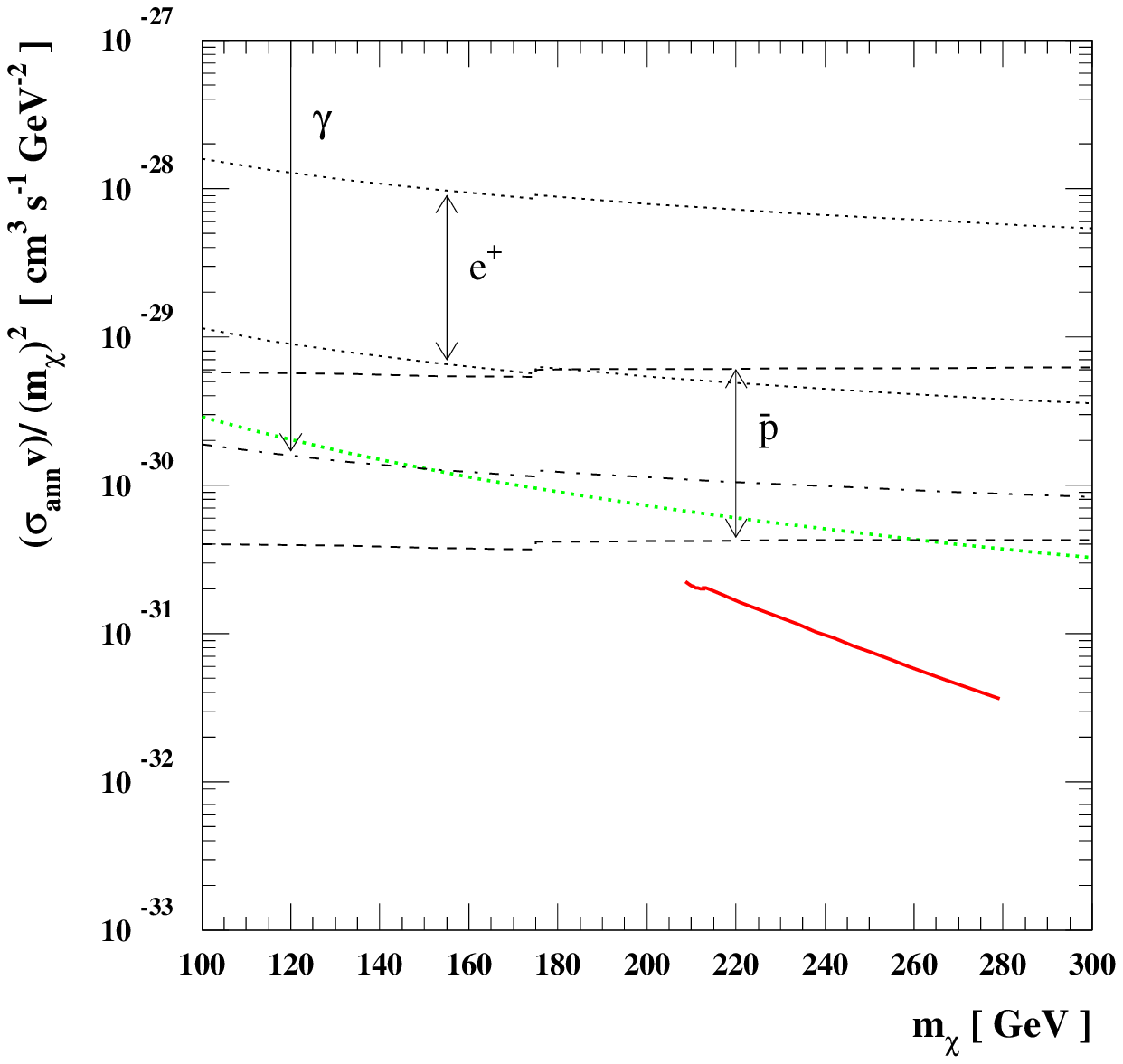,width=0.49\textwidth}
\epsfig{file=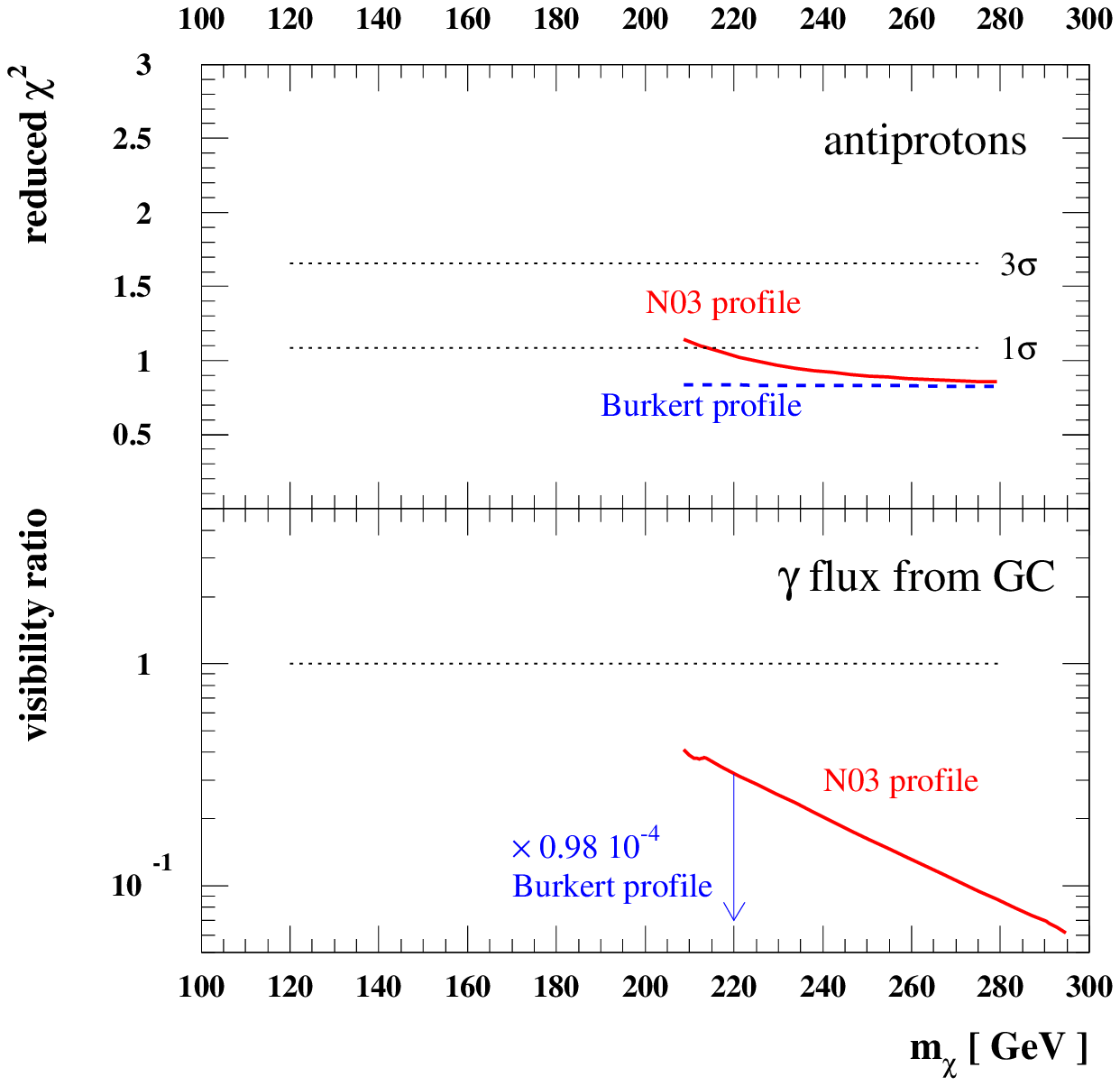,width=0.49\textwidth}}
\caption{In the left panel we show the annihilation cross section divided by the neutralino mass squared (red solid line) for our chosen stop coannihilation region. Also shown are the current limits coming from $\bar{p}$, $e^+$ and $\gamma$ measurements. The lower set of these curves are for the N03 halo profile and the upper for the Burkert profile. In the right panel, we show the reduced $\chi^2$ for the fits to antiproton data for these models and the visibility of gamma rays from annihilation at the galactic center.}
\label{fig:stopxsec}
\end{figure}

\begin{figure}
\centerline{\epsfig{file=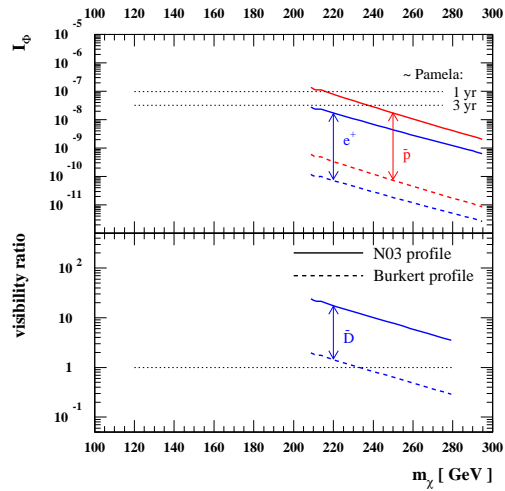,width=0.49\textwidth}}
\caption{Future detection prospects for our chosen stop coannihilation region. The lower set of curves are for the N03 profile and the upper for the Burkert profile. The direct detection scattering cross section is too low to be seen even by future detectors.}
\label{fig:stopcomp}
\end{figure}

This is the regime at small $m_{1/2}$, moderate $m_0$ and large $|A_0|$.
We focus here on a sample case in which the parameter $A_0$ is tuned to generate
a thin slice where neutralino-stop coannihilations drive the relic abundance into the
cosmologically interesting range: we choose $A_0 = -3000$ GeV, $\tan\beta = 5$ and ${\rm sign}(\mu)<0$. The 
stop coannihilation strip opens up at small $m_{1/2}$ and significantly larger $m_0$,
in a region where the neutralino is still a rather pure bino and  has a mass 
of about 100--300 GeV. In many cases, the LEP constraint on the $H_2^0$ boson mass severly affects these regions and can rule out the entire region with relic densities in our desired range. For the parameters we have chosen here, the Higgs constraint comes in at lower masses and sets a lower limit on the neutralino mass of 208 GeV.

In Fig.~\ref{fig:stopmuflux-sigma} we show the direct detection scattering cross section and the expected muon fluxes for these models together with current and expected future sensitivities. The dominating annihilation channel for these models is gluon-gluon below the $t \bar{t}$ threshold and $t \bar{t}$ above, i.e.\ a soft spectrum below $m_t$ and a hard above. We have used this to plot the expected sensitivity for IceCube \cite{icecube} (the jump at $m_t$ just reflects the change from soft to hard annihilation spectrum).
As is evident from these figures, none of these models can be detected with neutrino telescopes and direct detection experiments.

In Fig.~\ref{fig:stopxsec} we instead focus on signals from annihilations in the galactic halo. In this regime, neutralino pair annihilations are, above threshold, dominated by the $t \bar{t}$ channel, which is not S-wave suppressed and it is a rather copious source of antimatter particles. When $t \bar{t}$ is not kinematically allowed, below the $t \bar{t}$ threshold,  the gluon-gluon channel dominates, but this is never the case for the models displayed as the set of models below the threshold and beyond are excluded by the $H_2^0$ mass constraint from LEP. Annihilation to gluon-gluon below the threshold is though assumed in the exclusion curves shown. As can be seen in the left panel of the figure, we are now much closer to the current sensitivities, but still not quite to a level of excluding models. In the right panel, we check explicitly the corresponding reduced $\chi^2$ for antiproton fluxes against current data, which falls within the 1-$\sigma$ range for almost all models. We also compare with the gamma ray flux from the galactic center, showing that the predicted fluxes are always lower than the flux measured by EGRET for the GC source, even for the cuspy N03 profile.

We now move on to future detectors, and in Fig.~\ref{fig:stopcomp} we show the detection prospect by future searches. The antiprotons are marginally detectable with future detectors, like PAMELA in our cuspy N03 profile, whereas the positron fluxes are too low. The antideuteron fluxes, on the other hand, are high enough to be detectable by e.g.\ the GAPS probe, especially if the halo profile is cuspy, like the N03.

To summarize, the stop coannihilation region seems to be best probed by the antideuteron fluxes as measured by e.g.\ a future GAPS probe. The direct detection and neutrino telescope rates are far too low to be seen even by anticipated future detectors. The reason that annihilation in the halo is more advantageous, is probably that annihilation into $t \bar{t}$ dominates and this gives rather large fluxes of cosmic rays, whereas the scattering rates are low since the crossed diagram is not significant for the scattering cross section.

\subsection{Focus point region}

We describe now the regime at $m_0 \gg m_{1/2}$. Sitting on a isolevel curve at fixed
relic abundance in this regime essentially means that, at each mass we are selecting a preferred  Higgsino fraction for the lightest neutralino. At a fixed Higgsino fraction, the annihilation cross section into gauge bosons, which dominates for pure Higgsinos, 
decreases as the mass increases, hence the higher in mass we go, the more
pure Higgsinos we need to have to compensate for this decrease. In this respect,
$\tan\beta$ do not play much of a role, and the cases for positive and negative
$\mu$ are perfectly analogous (except that again the $\mu<0$ case is slightly disfavored 
by $(g-2)$). We choose to show, as sample cases, slices of the parameter space at $A_0=0$, $\mu>0$ and $\tan\beta = 30$ or $\tan\beta = 50$.

\begin{figure}[t]
\centerline{
\includegraphics[width=0.49\textwidth]{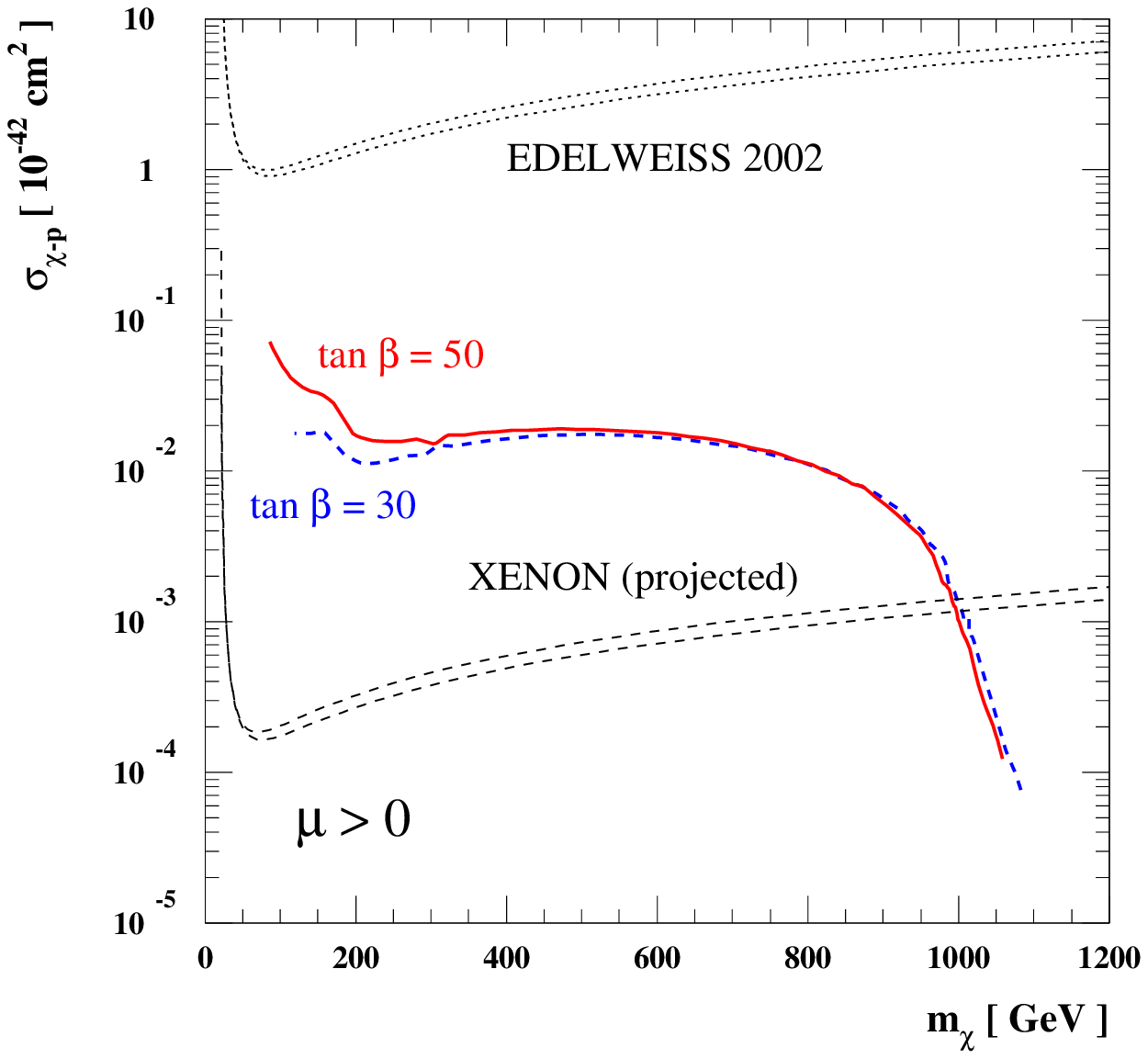}
\includegraphics[width=0.49\textwidth]{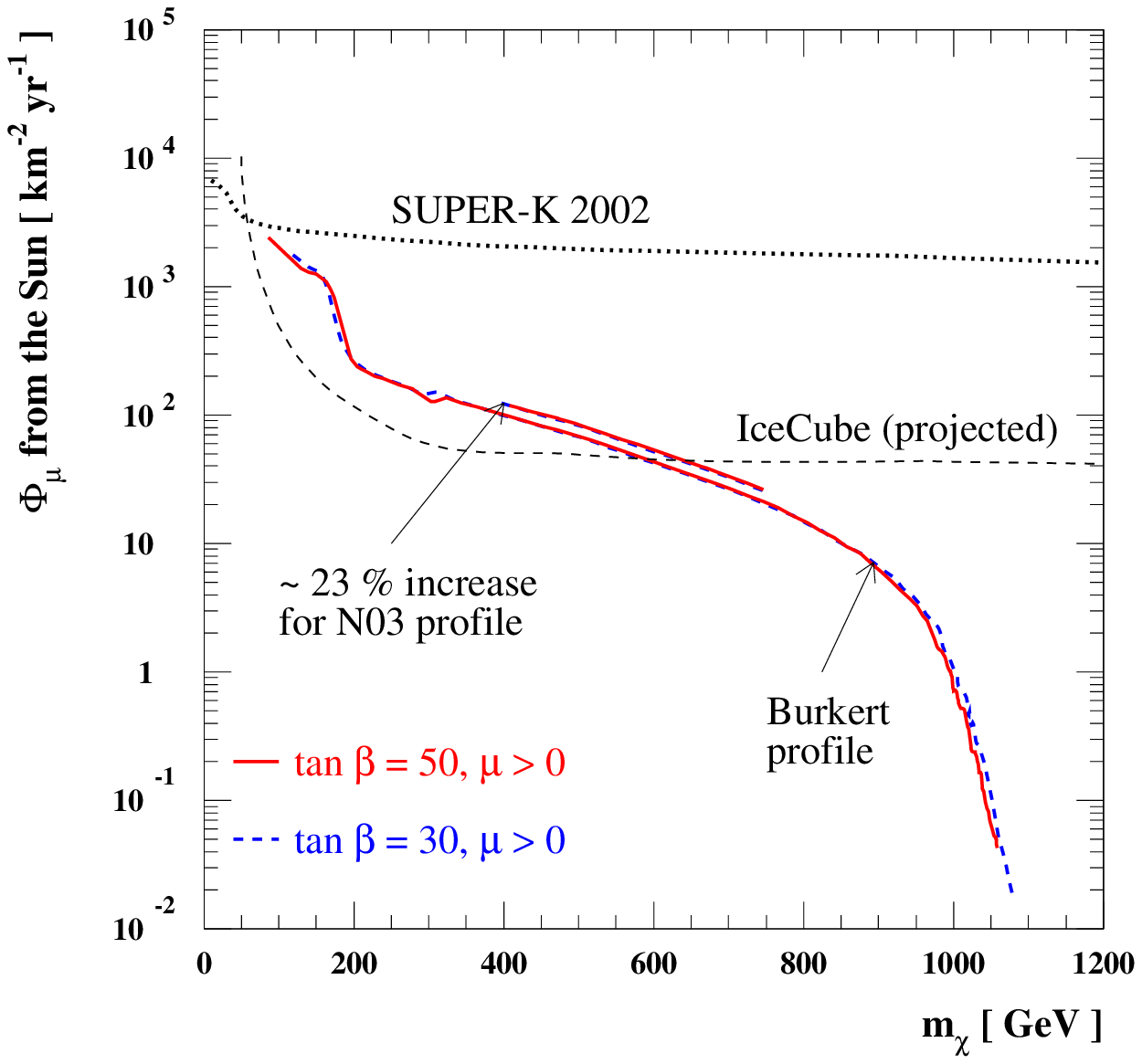}}
\caption{The expected direct detection spin independent scattering cross section (left panel) and neutrino-induced muon fluxes from the Sun (right panel) for two sample 
cases in the limit $m_0 \gg m_{1/2}$. Along each of the curves we have fixed $\Omega_\chi h^2 = 0.103$.
}
\label{foc1}
\end{figure}

The predicted direct detection spin independent scattering cross section and neutrino-induced muon fluxes from the Sun are shown in Fig.~\ref{foc1}. The spin independent scattering cross section is in this case dominated by one single diagram, the one mediated by $H^0_2$, as the squarks are extremely heavy, and $H^0_1$ turns out to be very heavy as well: it follows that, as can be seen, there is not much of a dependence on the $\tan\beta$ parameter, contrary to the cases at small $m_0$ we have discussed. Also, 
the cross sections are (relatively) large over most of the mass range, as there the Higgsino-Bino mixing in the lightest neutralino is fairly large, and we 
already pointed out a large mixing enhances the coupling of the neutralino to
CP-even Higgses.  Just at a mass of around 1~TeV we get to the limit of very pure Higgsinos
and the cross section at that point drops dramatically. Except for this corner, this
framework looks testable with the XENON detector (while unfortunately even this case is unconstrained by current experiments). For the expected sensitivity in IceCube, we have used the fact that for these models the neutralinos annihilate mainly to $Z^0 Z^0$, $W^+ W^-$ or $t \bar{t}$, i.e.\ they have hard annihilation spectra. The signal due to the neutrino flux from the Sun looks rather promising as well in a future perspective: the effect which is driving it
to levels which may be tested at future large size telescopes is the fact that the spin dependent scattering cross section is rather large, driven by the diagram with a $Z^0$
boson exchange (with Higgsino $Z^0$ coupling that does not depend on $\tan\beta$,
but again just on Higgsino fraction and mixing between the two Higgsino states, hence with the two cases at different $\tan\beta$ nearly overlapping). The two halo profiles change slightly the level of the flux, with about a 23\% increase for the N03 profile,
an effect which is mostly due to the higher local density (which increases accordingly the capture
rate), but with the two different velocity distribution functions playing some minor role.

\begin{figure}[t]
\centerline{
\includegraphics[width=0.49\textwidth]{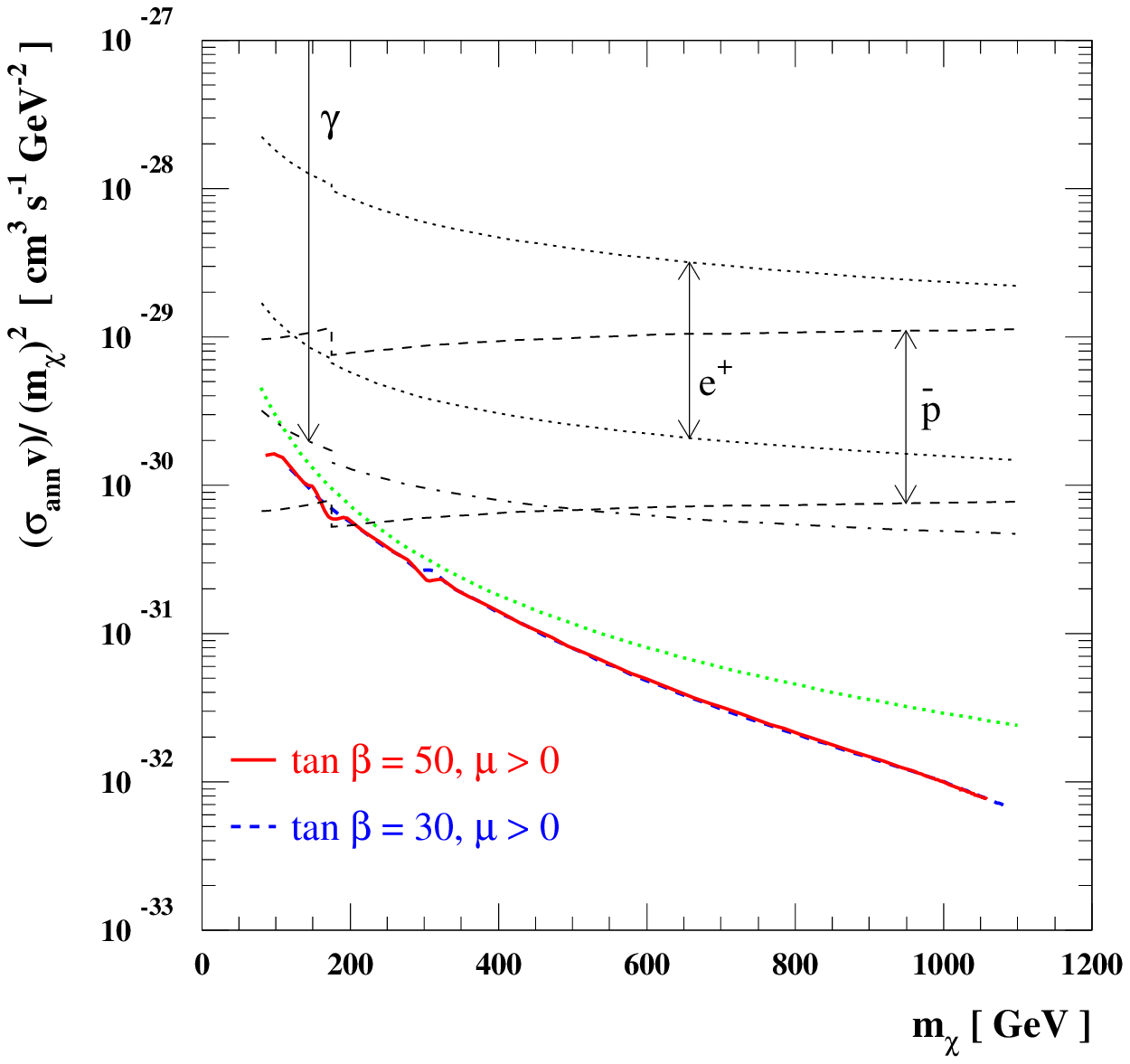}
\includegraphics[width=0.49\textwidth]{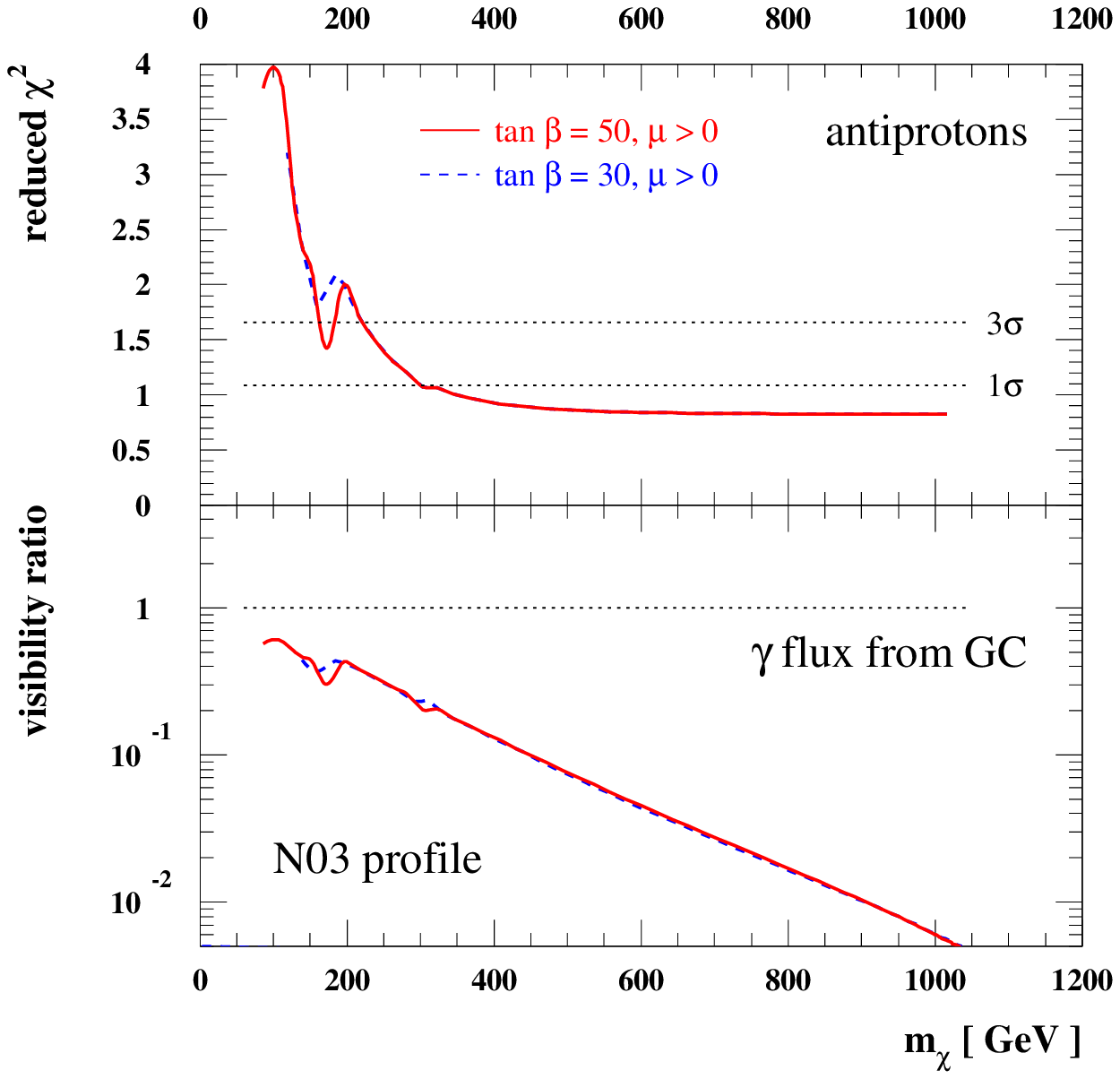}}
\caption{In the left panel we show the annihilation cross section divided by the neutralino mass squared for two sample cases in the focus point region. In the right panel we compare the predicted antiproton and gamma-ray fluxes against current data, see the text for details. In the latter we display just the case for the N03 profile, as for the Burkert profile signals are much smaller than measured fluxes.
}
\label{foc2}
\end{figure}

As shown in Fig.~\ref{foc2}, the annihilation cross section of neutralinos in halos today is  rather accurately set by the relic abundance, with curves at different $\tan\beta$ overlapping again and nearly aligning over the curve from which $\sigma_{ann} v$
is estimated using Eq.~(\ref{simpleomega}). Comparison with current data sets is performed in exactly the same way as for the other cases we discussed. Except for the
antiproton flux in case of low masses and for the N03 profiles, current data do not set constraints on the configurations we have chosen. 

\begin{figure}[t]
\centerline{
\includegraphics[width=0.49\textwidth]{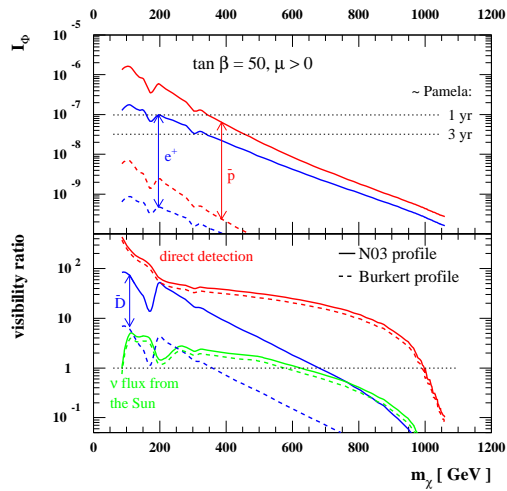}}
\caption{Future detection prospect in the focus point region, for the $\tan\beta = 50$ case we discussed in the text. The $\tan\beta = 30$ case is perfectly analogous.
}
\label{foc3}
\end{figure}

The prospects for detection of these dark matter candidates in the future are shown in 
Fig.~\ref{foc3}. The focus point region, and in particular the slice of it we have selected
at fixed value of the relic abundance and therefore with large Bino-Higgsino mixing, is the region which looks more promising from the point of view of dark matter detection.
Direct detection will test models up to very large masses, cross checks with very clean
signatures may come as well from neutrino telescopes, and the measurement of 
$\bar{D}$ flux (for small neutralino masses, even in case of the Burkert profile).
If the halo is cuspy, the neutralino-induced antiproton and positron flux may be at a detectable level as well, and might be soon singled out in upcoming experiments.

\section{Conclusion}

We have performed a detailed analysis of current limits and detection prospects of neutralino dark matter in the mSUGRA framework. We have focused on models with a thermal relic density, as estimated with the \ds\ numerical package, in the currently favored cosmological range, and considered all relevant regimes in the parameter space.   
Direct and indirect detection rates have been computed implementing two dark matter halos, with fully consistent density profiles and velocity distribution functions, and opposite histories for the transition between the stage of a CDM halo prior to the baryon infall and a halo embedded in a galaxy with inner portion dominated by the 
luminous components, as is the case for the Milky Way halo. This has allowed, for the first time, a fully consistent comparison between direct and indirect detection.

In general, we can conclude that most of the mSUGRA models considered here are not excluded by any of the current dark matter searches. For some models (low mass stau coannihilation region and low mass focus point region), we overproduce antiprotons and gamma rays from the galactic center in our cuspy N03 profile (but not with the cored Burkert profile). 

For future experiments, 
we have found that in the region of small $m_0$, direct detection is rather promising
if $\mu$ is positive and $\tan\beta$ is large, a feature due to the scattering amplitudes mediated by CP-even Higgs bosons summing coherently and to the coupling in the
$H^0_1\,d\,\bar{d}$ vertex becoming large. In the same region, but for different reasons,
the neutralino-induced antiproton, positron and especially
antideuteron fluxes could be detectable.
In the stop coannihilation region, both the direct detection and the neutrino telescope rates are too low to be detectable even with future experiments. The most promising technique to test these models is to search for an antideuteron flux with an experiment like GAPS; large fluxes follow in this case from large annihilation rates into top quarks.
Finally, in the funnel region, direct detection looks very promising because of the large
portion of both Bino and Higgsino in the lightest neutralino. An eventual signal in direct
detection experiments may be cross checked with the measurement of the induced neutrino flux from the Sun, and may even be anticipated through measurements 
of cosmic ray antimatter fluxes; both of these kinds of signals are expected to be large because of the large Higgsino fraction in this region.

As we have stressed, this analysis applies to one specific framework,
and relies on specific effects emerging in that framework. Extrapolations to other
perfectly viable models are possible, once the corresponding relevant effects are singled out in those models as well.

\section*{Acknowledgements} 

J.E.\ was supported by the Swedish Research Council.
P.U.\ was supported in part by the RTN project under 
grant HPRN-CT-2000-00152 and by the Italian INFN under the
project ``Fisica Astroparticellare''. P.U.\ would like to thank Stefano Profumo for stimulating discussions.


\end{document}